\documentclass[journal]{IEEEtran}
\ifCLASSINFOpdf
   \usepackage[pdftex]{graphicx}
   \graphicspath{{./figure/}}
   \DeclareGraphicsExtensions{.pdf,.jpeg,.png,.eps}
\else
   \usepackage[dvips]{graphicx}
   \graphicspath{{./figure/}}
   \DeclareGraphicsExtensions{.eps}
\fi

\usepackage{tikz,siunitx}
\usepackage{amsmath,amssymb}
\usepackage{cite}
\usepackage{graphicx,color}
\usepackage{subfigure}
\usepackage{url}
\usepackage{hyperref}

\newtheorem{thm}{Theorem}
\newtheorem{cor}{Corollary}[thm]

\newcommand{\mat}[1]{\mbox{\boldmath $#1$}}

\renewcommand{\eqref}[1]{(\ref{#1})}
\newcommand{\figref}[1]{Fig. \ref{#1}}
\definecolor{sblue}{RGB}{0,51,160}
\ifCLASSINFOpdf

\else

\fi

\begin{document}

\title{Hermite Expansion Model and LMMSE Analysis for Low-Resolution Quantized MIMO Detection}

\author{Lifu Liu, Yi Ma, and Na Yi
\thanks{The authors are with 5GIC and 6GIC, Institute for Communication Systems, 
University of Surrey, Guildford,  United Kingdom, GU2 7XH,  e-mail: (lifu.liu, y.ma, n.yi)@surrey.ac.uk. Tel: +44 1483 683609. 
(Corresponding author: Yi Ma)}
\thanks{This work was partially funded by the European Commission Horizon 2020 5G-DRIVE and partially by 5GIC and 6GIC research projects.
}
}

\markboth{}%
{Shell \MakeLowercase{\textit{et al.}}: Bare Demo of IEEEtran.cls for IEEE Journals}

\maketitle

\begin{abstract}
In this paper, the Hermite polynomials are employed to study linear approximation models of narrowband multi-antenna signal reception (i.e., MIMO)
with low-resolution quantizations. This study results in a novel linear approximation using the second-order Hermite expansion (SOHE).
The SOHE model is not based on those assumptions often used in existing linear approximations. Instead, the quantization distortion 
is characterized by the second-order Hermite kernel, and the signal term is characterized by the first-order Hermite kernel. It is shown that the 
SOHE model can explain almost all phenomena and characteristics observed so far in the low-resolution MIMO signal reception. When the SOHE 
model is employed to analyze the linear minimum-mean-square-error (LMMSE) channel equalizer, it is revealed that the current LMMSE algorithm 
can be enhanced by incorporating a symbol-level normalization mechanism. 
The performance of the enhanced LMMSE algorithm is demonstrated 
through computer simulations for narrowband MIMO systems in Rayleigh fading channels.
\end{abstract}


\IEEEpeerreviewmaketitle

\section{Introduction}\label{sec1}
\IEEEPARstart{M}{assive multiple-input multiple-output} (mMIMO) is a well-recognized radio frequency (RF) technology for highly 
spectrum-efficient communications \cite{6375940}. Current mMIMO technology has two main architectures, i.e., the 
analogue-digital hybrid architecture and the fully digital (FD) architecture (see \cite{1519678}). 
In the FD-mMIMO system, every antenna element is connected 
to a dedicated RF chain. It has been revealed that the energy consumption of each RF chain grows exponentially with the resolution of signal quantizers 
\cite{6457363, 761034}. This has recently motivated the use of low-resolution (mainly $1$-$3$ bit) quantizers for FD-mMIMO 
(e.g. \cite{5351659, DBLP:journals/corr/RisiPL14, 6891254}). The information-theoretic study of low-resolution quantized FD-mMIMO can be found in the literature 
(e.g. \cite{6891254, 7106472, 7080890}). 

In the scope of estimation and detection theory, digital signal processing problems such as channel estimation, synchronization and 
signal detection can be fundamentally changed due to the use of low-resolution quantizers \cite{6987288, 7088639, 9311778 }. This is because wireless systems  
become non-linear and non-Gaussian, and such violates the hypothesis of linear and Gaussian process that is commonly adopted in 
the conventional MIMO systems. Specifically for the signal detection problem, the maximum-likelihood detection becomes even more 
complicated as it is no longer equivalent to the integer least-squares problem \cite{4475570,9145094}. This is particularly true for FD-mMIMO 
systems with $1$-bit quantizers. 
In order to improve the computational efficiency, a number of near-maximum-likelihood algorithms have been reported in the literature (e.g. \cite{7439790, 8240630, 8345169}). However, they are still too complex to implement in practice. 
Approximate message passing (AMP) algorithms could offer near Bayesian-optimum solutions with much lower computational complexities (e.g. \cite{7355388, 7426735, 8234637}). 
Nevertheless, linear algorithms are practically more appealing for their lower complexities and simple architectures. 

The foundation of linear detection algorithms lies in a linear system model. Therefore, the central task of linear algorithm design is to find 
a good linear approximation of the non-linear communication system. In the literature, one of the widely used linear approximation models is 
the additive quantization noise model (AQNM) originally proposed in \cite{mezghani11}. It assumes the quantization distortion to be 
additive white Gaussian noise (AWGN) and correlated with the input of the quantizer. With the AQNM model, the LMMSE channel equalization and symbol-by-symbol detection algorithm has been extensively studied. 
Moreover, the AQNM model has been employed for the information-theoretic 
study of the achievable rate, capacity bounds or spectral efficiencies in \cite{7307134, 7308988, 7876856, 7896590, 7420605}. 
In \cite{Mezghani2012}, a modified-AQNM model has been proposed by making the quantization noise uncorrelated with the input signal 
through Wiener-Hopf filtering. This modified version renders the derivation of auto-covariances and cross-covariances involved in the 
LMMSE analysis much easier. When the input signal is white Gaussian, we show in Section \ref{sec2b2} that the modified-AQNM is equivalent 
to the original AQNM model. Using the Bussgang's theory\footnote{Please 
see \cite{Bussgang52} for the detail of Bussgang's theory.}, the modified-AQNM model has been further generalized. 
Specifically for the $1$-bit quantization, the quantization noise is actually far from the Gaussian assumption. Then, the Bussgang's theory 
has been used in \cite{Mezghani2012} to derive an exact form of the relevant auto-covariances and cross-covariances for the LMMSE 
channel equalizer. Other relevant works that use the Bussgang's theory for linear receiver design 
or performance analysis can be found in \cite{nguyen2019linear,7931630,7894211}.

The hypothesis of Gaussian quantization noise renders the AQNM model and its variations not sufficiently accurate for some 
cases (see the detailed discussion in \cite{8337813}).  Moreover, it has been observed that the AQNM-based LMMSE channel equalizer can introduce 
a scalar ambiguity in the signal amplitude. This scalar ambiguity is not a problem for constant-modulus modulations such as M-ary 
phase-shift-keying (M-PSK). However, it is detrimental to non-constant-modulus modulations such as M-ary quadrature-amplitude-modulation (M-QAM), 
and thus it must be appropriately handled for instance through the energy normalization \cite{7439790,nguyen2019linear,tsefunda}. 
After all, the major concern is that the inaccuracy of the AQNM models could disadvantage the receiver optimization as far as 
non-constant-modulus modulations are concerned \cite{9144509,7247358}. Arguably, the generalized-AQNM model does take into account the scaling 
ambiguities. However, we find the current studies rather intuitive, and that a more rigorous analytical study is needed to develop a deeper 
understanding of the quantization distortion as well as its impact on the LMMSE channel equalizer. This forms the major motivation of our work. 

The major contribution of our work lies in the employment of Hermite polynomials to develop the aforementioned deeper understanding. 
This study results in a novel linear approximation model using the second-order Hermite expansion (SOHE). 
In brief, the SOHE model can be described by the following vector form (see the detail in Section \ref{sec3})
\begin{equation}\label{eqn001}
\mathbf{y}=\mathcal{Q}_b(\mathbf{r})\approx\lambda_b\mathbf{r}+\mathbf{q}_b,
\end{equation}
where $\mathcal{Q}_b(\cdot)$ is the $b$-bit uniform quantizer introduced in \cite{Proakis2007}, 
$\mathbf{r}, \mathbf{y}\in\mathbb{C}^{K\times 1}$ are the input and output of the quantizer, respectively, 
$\lambda_b$ is the coefficient of the first-order Hermite kernel which is a function of the resolution of the quantizer ($b$), 
and $\mathbf{q}_b\in\mathbb{C}^{K\times 1}$ is the quantization distortion with its characteristics related to the resolution of the 
quantizer ($K$: the size of relevant vectors). The SOHE model differs from the existing AQNM models mainly in two folds: 

{\em 1)} The Hermite coefficient ($\lambda_b$) in the SOHE model describes how the signal energy changes with respect to the 
resolution of the quantizer. The relationship between $\lambda_b$ and the resolution $b$ is mathematically formulated, 
based on which the characteristics of $\lambda_b$ are exhibited through our analytical work. 

{\em 2)} The quantization distortion ($\mathbf{q}_b$) is modeled as the second-order Hermite polynomial of the input signal 
($\mathbf{r}$). There is no imposed assumption for the quantization distortion to be white Gaussian as well as their correlation behavior 
with the input signal.  It will be shown in Section \ref{sec3}, through mathematical analysis, that the cross-correlation between $\mathbf{q}_b$ 
and the input $\mathbf{r}$ depends on the stochastic behavior of the input signal. When the input is an independent white Gaussian process, 
the quantization distortion can be considered to be uncorrelated with the input signal. 

With the above distinctive features, we find that the SOHE model can be used to explain almost all interesting phenomena observed so far 
in the research of low-resolution quantized MIMO signal detection. When using the SOHE model for the LMMSE analysis, our analytical work shows 
that the current LMMSE algorithm should be enhanced by incorporating a symbol-level normalization mechanism, and thereby resulting in an 
enhanced-LMMSE (e-LMMSE) channel equalizer. The performance gain of e-LMMSE is demonstrated through extensive computer simulations in 
Rayleigh fading channels. 

In addition, as response to the reviewers' comments, we enrich our technical contribution with the SOHE-based LMMSE channel estimation approach. 
It is found that the SOHE-LMMSE channel estimator can offer comparable sum spectral efficiency (SE) with the state-of-the-art (SOTA) because the performance is limited by the channel estimation error.

The rest of this paper is organized as follows. Section II presents the system model, preliminaries and problem statement. 
Section III presents the Hermite expansion model. Section IV presents the LMMSE analysis. Section V presents the simulation results, 
and finally Section VI draws the conclusion. 

\subsubsection*{Notations}
Regular letter, lower-case bold letter, and capital bold letter represent scalar, vector, and matrix, respectively. 
$\Re(\cdot)$ and $\Im(\cdot)$ represent the real and imaginary parts of a complex number, respectively. 
The notations $[\cdot]^T$, $[\cdot]^H$, $[\cdot]^*$, $[\cdot]^{-1}$, $\left \| \cdot \right \|$, $\mathrm{trace}(\cdot)$ and 
$\mathbb{D}(\cdot)$ represent the transpose, Hermitian, conjugate, inverse, Euclidean norm, trace and a matrix formed by the diagonal of a matrix 
(a vector or a scalar if appropriate), respectively. $\mathbb{E}\left [ \cdot \right ]$ denotes the expectation, $\mathbf{I}$ denotes the identity matrix,
and $\otimes$ denotes the Kronecker product.

\section{System Model, Preliminaries and\\ Problem Statement}\label{sec2}
This section introduces the mathematical model of the uplink MIMO signal reception with low-resolution quantizers. This is then followed by 
a review of current linear approximation models as well as their related LMMSE channel equalizers. This review is important in the sense that it can 
help to understand the SOTA as well as their differences from the SOHE model. It is perhaps worth noting that we do not put an 
emphasis on the mMIMO system mainly to keep our work as generic as possible. 

\subsection{System Model}\label{sec2a}
Similar to many other works in the SOTA analysis (e.g. \cite{7307134, 7876856, 7896590, 7420605}), we also consider a narrowband FD-mMIMO network, where a set of single-antenna transmitters $(N)$ simultaneously send their messages to 
a receiver having a large number of receive antennas $(K)$. Denote $s_n$ to be the information-bearing symbol sent by the $n^\mathrm{th}$ transmitter ($n=0,...,N-1$). It is commonly assumed that $s_n$ is drawn from a finite alphabet-set with equal probability and fulfills: 
$\mathbb{E}(s_n)=0$, $\mathbb{E}(s_ns_n^*)=1$, $\mathbb{E}(s_ns_m^*)=0$, $_{\forall n\neq m}$.
With the ideal quantization, the received discrete-time signal at the baseband ($\mathbf{r}$) is expressible as
\begin{equation}\label{eqn002}
	\mathbf{r}=\sum_{n=0}^{N-1}\mathbf{h}_ns_n+\mathbf{v},
\end{equation}
where $\mathbf{h}_n\in\mathbb{C}^{K\times1}$ is the channel vector corresponding to the $n^\mathrm{th}$ transmitter to the receiver link, 
and $\mathbf{v}\in\mathbb{C}^{K\times1}$ is the white Gaussian thermal noise with zero mean and auto-covariance $N_0\mathbf{I}$. 
Define $\mathbf{H}\triangleq[\mathbf{h}_0, ..., \mathbf{h}_{N-1}]$ and 
$\mathbf{s}\triangleq[s_0,...,s_{N-1}]^T$. The linear model \eqref{eqn002} can be rewritten into the following matrix form
\begin{equation}\label{eqn003}
\mathbf{r}=\mathbf{H}\mathbf{s}+\mathbf{v}.
\end{equation}
Feeding $\mathbf{r}$ into the $b$-bit low-resolution quantizer results in
\begin{equation}\label{eqn004}
\mathbf{y}=\mathcal{Q}_b(\Re(\mathbf{r}))+j\mathcal{Q}_b(\Im(\mathbf{r})),
\end{equation}
where the quantization is individually performed in the real and imaginary domains. 

To reconstruct the signal block $\mathbf{s}$ at the receiver (i.e., the signal detection), the channel knowledge $\mathbf{H}$ is usually assumed 
in the literature (e.g. \cite{5592653, 8320852, 8610159, 7155570}). There are also quite a few published works discussing 
about the channel estimation as well as the signal 
reconstruction based upon various channel knowledge imperfections (e.g. \cite{7439790, 7355388, 7247358, 5501995, 708938}). 
Those are indeed very interesting research issues. However, 
in order to make our work well focused on the signal reconstruction, we assume the availability of $\mathbf{H}$ throughout the paper 
and describe the signal reconstruction procedure as the following input-output relationship
\begin{equation}\label{eqn005}
	\hat{\mathbf{s}}=g(\mathbf{y}, \mathbf{H}),
\end{equation}
where $\hat{\mathbf{s}}$ is the reconstructed version of $\mathbf{s}$. In the following contents, our discussion will be focused on 
the linear approximation models and LMMSE analysis. Optimum and near-optimum approaches are not relevant to our discussion 
and therefore skipped. 

\subsection{Linear Approximation Models and LMMSE Analysis}\label{sec2b}
Our SOTA analysis shows that there are mainly three linear models to approximate the non-linear model \eqref{eqn004}, and they can 
lead to different LMMSE formulas. 
\subsubsection{The AQNM Model}\label{sec2b1}
this model can be mathematically described by (see \cite{mezghani11, 5351659})
\begin{equation}\label{eqn006}
	\mathbf{y}\approx\mathbf{z}_A\triangleq\mathbf{r}+\mathbf{q}_A.
\end{equation}
There are two assumptions for the AQNM model:
\begin{itemize}
\item[A1)] The quantization distortion $\mathbf{q}_A$ is AWGN;
\item[A2)] $\mathbf{q}_A$ is correlated with the input signal $\mathbf{r}$.
\end{itemize}

With this linear approximation model, the LMMSE channel equalizer ($\mathbf{G}^\star$) can be obtained by solving the following MMSE objective function
\begin{IEEEeqnarray}{ll}
	\mathbf{G}^\star&=\underset{\mathbf{G}}{\arg\min}~\mathbb{E}\|\mathbf{s}-\mathbf{G}\mathbf{y}\|^2,\label{eqn007}\\
	&\approx\underset{\mathbf{G}}{\arg\min}~\mathbb{E}\|\mathbf{s}-\mathbf{G}\mathbf{z}_A\|^2\label{eqn008}.
\end{IEEEeqnarray}
The solution to \eqref{eqn008} is provided in \cite{mezghani11}, i.e.,
\begin{equation}\label{eqn009}
	\mathbf{G}^\star=\mathbf{H}^H(N_0\mathbf{I}+\mathbf{HH}^H+\mathrm{nondiag}(\rho_b\mathbf{HH}^H))^{-1},
\end{equation}
where $\rho_b$ is the distortion factor indicating the relative amount of quantization noise
generated, and it is a function of $b$; see the specific discussion in the relevant literature \cite{1057548, 6891254, 7106472}.

\subsubsection{The Modified-AQNM Model}\label{sec2b2}
the mathematical form of this linear model is given by (see \cite{Mezghani2012}):
\begin{equation}\label{eqn010}
	\mathbf{y}\approx\mathbf{z}_B\triangleq\mathbf{C}_{yr}\mathbf{C}_{rr}^{-1}\mathbf{r}+\mathbf{q}_B,
\end{equation}
where $\mathbf{C}_{yr}$ is the cross-covariance matrix between $\mathbf{y}$ and $\mathbf{r}$, $\mathbf{C}_{rr}$ is the 
auto-covariance matrix of $\mathbf{r}$, and $\mathbf{q}_B$ is the quantization distortion. Different from the AQNM model in \eqref{eqn006},
the assumption here is:
\begin{itemize}
\item[A3)] the quantization distortion $(\mathbf{q}_B)$ is uncorrelated with the input signal $\mathbf{r}$.
\end{itemize}
Moreover, the condition A1) is not always assumed. 

Define $\overline{\mathbf{H}}\triangleq\mathbf{C}_{yr}\mathbf{C}_{rr}^{-1}\mathbf{H}$ and 
$\mat{\varepsilon}\triangleq\mathbf{C}_{yr}\mathbf{C}_{rr}^{-1}\mathbf{v}+\mathbf{q}_B$. The modified-AQNM model \eqref{eqn010} 
can be represented by the following canonical form  
\begin{equation}\label{eqn011}
\mathbf{z}_B=\overline{\mathbf{H}}\mathbf{s}+\mat{\varepsilon}.
\end{equation}
The auto-covariance matrix of $\mat{\varepsilon}$ is given by \cite[(9)]{Mezghani2012},
\begin{equation}\label{eqn012}
	\mathbf{C}_{\varepsilon\varepsilon}=\mathbf{C}_{yy}-\mathbf{C}_{yr}\mathbf{C}_{rr}^{-1}\mathbf{C}_{ry}
	+N_0\mathbf{C}_{yr}\mathbf{C}_{rr}^{-1}\mathbf{C}_{rr}^{-1}\mathbf{C}_{ry}.
\end{equation}
This is however too complex for the LMMSE analysis. 

Applying the assumption A1) onto the quantization distortion $\mathbf{q}_B$, it has been shown that the following approximation of 
covariance matrices applies
\begin{equation}\label{eqn013}
	\mathbf{C}_{ry}\approx(1-\rho_b)\mathbf{C}_{rr}\approx\mathbf{C}_{yr},
\end{equation}
\begin{equation}\label{eqn014}
	\mathbf{C}_{\varepsilon\varepsilon}\approx(1-\rho_b)^2N_0\mathbf{I}+(1-\rho_b)\rho_b\mathbb{D}(\mathbf{C}_{rr}).
\end{equation}
Applying \eqref{eqn013} into \eqref{eqn011} results in
\begin{equation}\label{eqn015}
	\mathbf{z}_B\approx(1-\rho_b)\mathbf{H}\mathbf{s}+\mat{\varepsilon}.
\end{equation}
Then, the LMMSE objective function reads as
\begin{equation}\label{eqn016}
	\mathbf{G}^\star=\underset{\mathbf{G}}{\arg\min}~\mathbb{E}\|\mathbf{s}-\mathbf{G}\mathbf{z}_B\|^2.
\end{equation}
Solving \eqref{eqn016} results in
\begin{equation}\label{eqn017}
	\mathbf{G}^\star=(1-\rho_b)^{-1}\mathbf{H}^H\Big(\mathbf{C}_{rr}+\frac{\rho_b}{1-\rho_b}\mathbb{D}(\mathbf{C}_{rr})
	\Big)^{-1},
\end{equation}
where $\mathbf{C}_{rr}=\mathbf{HH}^H+N_0\mathbf{I}$. This equation seems to be different from \eqref{eqn009}. However, if 
we incorporate the term $(1-\rho_b)^{-1}$ into the auto-covariance term inside the bracket, \eqref{eqn017} immediately turns 
into \eqref{eqn009}. Arguably, we can consider the linear approximations \eqref{eqn006} and \eqref{eqn010} to be equivalent 
when the assumption A1) is adopted. 

\subsubsection{The Generalized-AQNM Model}\label{sec2b3}
the modified-AQNM model can be extended to the following generalized version (see \cite{Mezghani2012})
\begin{equation}\label{eqn018}
\mathbf{z}_C=\mathbf{\Lambda}_b\mathbf{r}+\mathbf{q}_C,
\end{equation}
where the quantization distortion $\mathbf{q}_C$ is assumed to be uncorrelated with $\mathbf{r}$ (i.e., the assumption A3), 
and $\mathbf{\Lambda}_b$ is a diagonal matrix with its characteristics related to the low-resolution quantizer. 

Consider the quantizer $y=\mathcal{Q}_b(x),~x\in(-\infty, \infty)$, to be a stair function with its input range being divided into $M=2^b$ 
sub-ranges \footnote{
 The dynamic of sub-ranges is promised by the automatic gain control (AGC), which aims at keeping the amplitude of the output signal $y$ substantially constant or to vary only within a small range \cite{664234, 1092057}. In order to focus on the analysis of the low-resolution quantization process, the ideal AGC is assumed in this paper.} .
Define $(\tau_m, \tau_{m+1})$ to be the $m^\mathrm{th}$ sub-range. The quantizer can be represented by
\begin{equation}\label{eqn019}
\mathcal{Q}_b(x)=x_m,~x\in(\tau_m, \tau_{m+1}),~_{m=0, ..., M-1,}
\end{equation}
where in general $x_m$ can be an appropriately chosen value within the range of $(\tau_m, \tau_{m+1})$ depending on the design 
specification \cite{1057548,Liu2021vtc}; and $\tau_0=-\infty$, $\tau_{M}=\infty$. Then, the diagonal matrix $\mathbf{\Lambda}_b$ is 
expressed by
\begin{IEEEeqnarray}{ll}\label{eqn020}
\mathbf{\Lambda}_b&=\mathbb{D}(\mathbf{C}_{rr})^{-\frac{1}{2}}
\sum_{m=0}^{M-1}\frac{x_m}{\sqrt{\pi}}\Big(
\exp(-\tau_m^2\mathbb{D}(\mathbf{C}_{rr})^{-1})\nonumber\\
&\quad\quad\quad\quad\quad\quad\quad\quad-\exp(-\tau_{m+1}^2\mathbb{D}(\mathbf{C}_{rr})^{-1})\Big).
\end{IEEEeqnarray}
Generally, the analysis of $\mathbf{C}_{\varepsilon\varepsilon}$ is highly complex, and it does not result in a closed-form solution. 
Specifically for the special case of symmetric $1$-bit quantization, the assumption of Gaussian quantization noise is not suitable. 
Using the Bussgang's theorem, the approximations \eqref{eqn013}-\eqref{eqn014} can now be replaced by the 
exact forms (a slightly alternated version from \cite{Mezghani2012, nguyen2019linear})
\begin{equation}\label{eqn021}
	\mathbf{C}_{yr}=\sqrt{\frac{2}{\pi}}\mathbb{D}(\mathbf{C}_{rr})^{-\frac{1}{2}}\mathbf{C}_{rr},
\end{equation}
\begin{IEEEeqnarray}{ll}\label{eqn022}
	\mathbf{C}_{\varepsilon\varepsilon}=&\frac{2}{\pi}\Big[\arcsin\Big(\mathbb{D}(\mathbf{C}_{rr})^{-\frac{1}{2}}\mathbf{C}_{rr}\mathbb{D}(\mathbf{C}_{rr})^{-\frac{1}{2}}\Big)-\nonumber\\
	&\mathbb{D}(\mathbf{C}_{rr})^{-\frac{1}{2}}\mathbf{C}_{rr}\mathbb{D}(\mathbf{C}_{rr})^{-\frac{1}{2}}+
	N_0\mathbb{D}(\mathbf{C}_{rr})^{-1}\Big].
\end{IEEEeqnarray}
Applying the above results for the LMMSE analysis leads to
\begin{equation}\label{eqn023}
	\mathbf{z}_B=\sqrt{\frac{2}{\pi}}\mathbb{D}(\mathbf{C}_{rr})^{-\frac{1}{2}}\mathbf{H}\mathbf{s}+\mat{\varepsilon}.
\end{equation}
With \eqref{eqn018}-\eqref{eqn020}, it is rather trivial to obtain the following form of LMMSE
\begin{equation}\label{eqn024}
	\mathbf{G}^\star=\sqrt{\frac{2}{\pi}}\mathbb{D}(\mathbf{C}_{rr})^{-\frac{1}{2}}\mathbf{H}
	\Big(\mathbf{C}_{\varepsilon\varepsilon}+\frac{2}{\pi}\mathbb{D}(\mathbf{C}_{rr})^{-1}\mathbf{HH}^H\Big)^{-1}.
\end{equation}
Here, we emphasize that \eqref{eqn024} holds only for the $1$-bit quantizer. 

\subsection{Statement of The Research Problem}
Section \ref{sec2b} shows already intensive research and appealing contributions on the linear approximation models as well as their relevant LMMSE analaysis. 
Nevertheless, there is still a need for a more extensive and rigorous study on this issue, which can make the linear approximation 
research more comprehensive and accurate. Moreover, a more comprehensive study could help to develop novel understanding of the behavior of 
LMMSE channel equalizer in the context of low-resolution MIMO signal reception. The following sections are therefore motivated.

\section{Hermite Polynomial Expansion for Linear Approximation}\label{sec3}
This section presents the Hermite polynomial expansion of the low-resolution quantization function as well as key characteristics of the 
SOHE model.

\subsection {Hermite Polynomial Expansion and The SOHE Model}
We start from the Laplace's Hermite polynomial expansion (see the definition in \cite[Chapter 22]{Poularikas_1999}) which is employed to 
represent the quantization function $y=\mathcal{Q}_b(x),~x\in(-\infty, \infty)$. The Hermite transform of $\mathcal{Q}_b(x)$ is given by (see 
\cite{60086})
\begin{equation}\label{eqn025}
\omega_l=\frac{1}{\sqrt{\pi}2^ll!}\int_{-\infty}^{\infty}\mathcal{Q}_b(x)\exp(-x^2)\beta_l(x)\mathrm{d}x,
\end{equation}
where $\beta_l(x)$ is the Rodrigues' formula specified by
\begin{equation}\label{eqn026}
\beta_l(x)=(-1)^l\exp(x^2)\Big[\frac{\partial^l}{\partial x^l}\exp(-x^2)\Big].
\end{equation}
With this result,  the Hermite polynomial expansion of $\mathcal{Q}_b(x)$ is given by
\begin{equation}\label{eqn027}
\mathcal{Q}_b(x)=\lim_{L\rightarrow\infty}\sum_{l=1}^{L}\omega_l\beta_l(x).
\end{equation}
The expression of $\omega_l$ can be simplified by plugging \eqref{eqn026} into \eqref{eqn025}, i.e.,  
\begin{equation}\label{eqn028}
\omega_l=\frac{(-1)^l}{\sqrt{\pi}2^ll!}\int_{-\infty}^{\infty}\mathcal{Q}_b(x)\Big[\frac{\partial^l}{\partial x^l}\exp(-x^2)\Big]\mathrm{d}x.
\end{equation}
Applying \eqref{eqn019} into \eqref{eqn028} results in
\begin{equation}\label{eqn029}
w_l=\frac{(-1)^l}{\sqrt{\pi}2^ll!}\sum_{m=0}^{M-1}x_m\int_{\tau_m}^{\tau_{m+1}}
\Big[\frac{\partial^l}{\partial x^l}\exp(-x^2)\Big]\mathrm{d}x.
\end{equation}

The SOHE model is based on the second-order Hermite expansion as below (i.e., $L=2$ in \eqref{eqn027})
\begin{IEEEeqnarray}{ll}\label{eqn030}
\mathcal{Q}_b(x)&=\sum_{l=1}^{2}w_l\beta_l(x)+O(w_3\beta_3(x)),\\
&=\lambda_bx+q_b(x),\label{eqn031}
\end{IEEEeqnarray}
where $\lambda_b$ is the coefficient corresponding to the first-order Hermite kernel, and $q_b$ is the second-order 
approximation of the quantization noise. Their mathematical forms are specified by
\begin{equation}\label{eqn032}
\lambda_b=2\omega_1,
\end{equation}
\begin{equation}\label{eqn033}
q_b(x)=4\omega_2x^2-2\omega_2+O(\omega_3\beta_3(x)).
\end{equation}
The derivation from \eqref{eqn030} to \eqref{eqn031} is by means of computing \eqref{eqn026} for $l=1,2$. 
The mathematical work is rather trivial and thus omitted. 

{\em Remark 1:}
The SOHE model in \eqref{eqn031} is certainly not accurate enough to capture the exact characteristics of the low-resolution quantizer. 
This is also true for all other existing linear approximation models. An accurate Hermite model often requires $L=100$ or more, and this is however 
too complex for an analytical study. Nevertheless, we will show that the SOHE model can already reflect key characteristics of the low-resolution 
quantizer. 

\subsection{The Scalar-SOHE Model and Characteristics}\label{3b}
The SOHE model is a linear approximation of the low-resolution quantizer, and thus it is not very different from other existing linear models 
if solely based on their mathematical forms. On the other hand, the key parameters of SOHE (i.e., $\lambda_b$ and $q_b(x)$) show 
different characteristics from the others. 

{\em 1)} Characteristics of the Hermite coefficient $\lambda_b$ can be summarized by the following statement. 
\begin{thm}\label{thm01}
Consider the case of symmetric $b$-bit  quantization with the following setup in \eqref{eqn029}
\begin{equation}\label{eqn034}
x_m=\left\{\begin{array}{ll}
\tau_{m+1},&\tau_{m+1}>0\\
\tau_m,&\tau_{m+1}<0
\end{array}\right.
\end{equation}
The Hermite coefficient $\lambda_b$ 
has the following properties:
\begin{equation}\label{eqn035}
\lambda_b\geq 1~\mathrm{and} \lim_{b\rightarrow\infty}\lambda_b=1.
\end{equation}
\end{thm}
\begin{IEEEproof}
See Appendix \ref{A}.
\end{IEEEproof}

With the ideal AGC, we assume that the input and output signals can be optimally scaled to meet the quantization boundaries.
{\em Theorem \ref{thm01}} provides two implications: {\em 1)} low-resolution quantizers can introduce a scalar ambiguity $\lambda_b$, 
which often amplifies the input signal in the digital domain. The principle on how the signal is amplified is analytically explained in 
Appendix \ref{A}; {\em 2)} In the SOHE model, the scalar ambiguity vanishes with the increase of resolution ($b$ or $M$). This is in line 
with the phenomenon that can be observed in reality. In other words, the SOHE model as well as the proof in Appendix \ref{A} can well 
explain the phenomenon of scalar ambiguity occurred in our practice. 

{\em 2)} Unlike other linear approximation models, the SOHE model does not impose the assumptions A1) and A2) (see Section \ref{sec2b})
onto the quantization noise $q_b$. Instead, $q_b$ is described as a function of the input signal $x$, with their statistical behaviors being 
analytical studied here. 
\begin{thm}\label{thm02}
 Suppose: C1) the input signal $x$ follows $\mathbb{E}(x)=0$. The cross-correlation between $x$ and $q_b$ depends on the third-order central moments of $x$. 
When the input signal $(x)$ is AWGN, the quantization noise can be considered to be uncorrelated with the input signal. Moreover, for the 
case of $b\rightarrow\infty$, the following result holds
\begin{equation}\label{eqn036}
\lim_{b\rightarrow\infty}q_b(x)=0.
\end{equation}
\end{thm}
\begin{IEEEproof}
See Appendix \ref{B}.
\end{IEEEproof}

The implication of {\em Theorem \ref{thm02}} lies in two folds: {\em 1)} the quantization noise cannot be easily assumed to be uncorrelated with the input signal. {\em Theorem \ref{thm02}} provides sufficient conditions for the hypothesis of uncorrelated quantization noise; {\em 2)} due to the use of second-order expansion for quantization function, it is possible that the SOHE-based quantization noise cannot well represent the characteristics of ideal quantization like \eqref{eqn036}. However, {\em Theorem \ref{thm02}} confirms that with the increasing of resolutions, the quantization noise which is a function of the input signal would approximate to zero. 

{\em Remark 2:} 
It is worthwhile to note that, for complex-valued signals, the quantization process is applied individually in the real and imaginary domains. 
Therefore, {\em Theorems \ref{thm01}-\ref{thm02}} apply straightforwardly to the complex-valued input signal.

\subsection{The Vector-SOHE Model and Characteristics}
The vector representation of the SOHE model has no fundamental difference from the scalar-SOHE model presented 
in \eqref{eqn031}. It can be obtained by applying \eqref{eqn031} into \eqref{eqn004}
\begin{IEEEeqnarray}{ll}\label{eqn037}
\mathbf{y}&=\lambda_b\mathbf{r}+\mathbf{q}_b,\\
&=\lambda_b\mathbf{H}\mathbf{s}+\underbrace{\lambda_b\mathbf{v}+\mathbf{q}_b}_{\triangleq\mat{\varepsilon}_b}.\label{eqn038}
\end{IEEEeqnarray}
The vector form of the quantization noise is specified by
\begin{equation}\label{eqn039}
\mathbf{q}_b=4\omega_2\Big(\Re(\mathbf{r})^2+j\Im(\mathbf{r})^2\Big)-2\omega_2,
\end{equation}
where $\Re(\mathbf{r})^2$ or $\Im(\mathbf{r})^2$ denotes the corresponding real-vector with the Hadamard power of $2$. 
With {\em Theorem \ref{thm02}}, we can reach the following conclusion about the vector-SOHE model. 

\begin{cor}\label{cor1}
Suppose that C2) each element of $\mathbf{H}$ is independently 
generated;  and C3) the number of transmit antennas ($N$) is sufficiently large. The following cross-covariance matrix can be obtained
\begin{equation}\label{eqn040}
\mathbf{C}_{qv}=\mathbb{E}(\mathbf{q}_b\mathbf{v}^H)=\mathbf{0}.
\end{equation}
\end{cor}
\begin{IEEEproof}
The condition C2) ensures that each element of the vector $[\mathbf{Hs}]$ is a sum of $N$ independently generated random variables. 
With the condition C3), the central limit theorem tells us that each element of $[\mathbf{Hs}]$  is
asymptotically AWGN. Since the thermal noise $\mathbf{v}$ is AWGN and independent from $[\mathbf{Hs}]$, 
the received signal $\mathbf{r}$ is approximately AWGN. In this case, {\em Theorem \ref{thm02}} tells us
\begin{equation}\label{eqn041}
\mathbf{C}_{qr}=\mathbb{E}(\mathbf{q}_b\mathbf{r}^H)=\mathbf{0}.
\end{equation}
Plugging \eqref{eqn003} into \eqref{eqn041} results in
\begin{IEEEeqnarray}{ll}\label{eqn042}
\mathbf{C}_{qr}&=\mathbb{E}(\mathbf{q}_b(\mathbf{Hs}+\mathbf{v})^H),\\
&=\mathbb{E}(\mathbf{q}_b(\mathbf{Hs})^H)+\mathbf{C}_{qv}=\mathbf{0}.\label{eqn043}
\end{IEEEeqnarray}
Since $\mathbf{v}$ is independent from $[\mathbf{Hs}]$, the only case for \eqref{eqn043} to hold is that both cross-covariance terms are zero. 
\eqref{eqn040} is therefore proved. 
\end{IEEEproof}

\begin{cor}\label{cor2}
Given the conditions C2) and C3),  the  auto-covariance matrix of the quantization noise ($\mathbf{C}_{qq}$) has the following 
asymptotical form
\begin{equation}\label{eqn044}
\mathbf{C}_{qq}=4\omega_2^2\Big(4\sigma_r^4\mathbf{I}+(2\sigma_r^4-\sigma_r^2+1)(\mathbf{1}\otimes\mathbf{1}^T)\Big),
\end{equation}
where $\sigma_{r}^2$ denotes the variance of $r_k, _{\forall k}$ when $N\rightarrow\infty$.
\end{cor}
\begin{IEEEproof}
See Appendix \ref{C}.
\end{IEEEproof}
\begin{thm}\label{thm03}
Suppose that C4) the information-bearing symbols $s_n,  _{\forall n},$ have their third-order central moments fulfilling the condition:
$\mathbb{E}(\Re(s_n)^3)=0$; $\mathbb{E}(\Im(s_n)^3)=0$. Then, the following cross-covariance holds
\end{thm}
\begin{equation}\label{eqn045}
\mathbf{C}_{\varepsilon s}=\mathbb{E}(\mat{\varepsilon}_b\mathbf{s}^H)=\mathbf{0}.
\end{equation}
\begin{IEEEproof}
The cross-covariance in \eqref{eqn045} can be computed as follows
\begin{IEEEeqnarray}{ll}
\mathbf{C}_{\varepsilon s}&=\mathbb{E}((\lambda_b\mathbf{v}+\mathbf{q}_b)\mathbf{s}^H),\label{eqn046}\\
&=\lambda_b\mathbf{C}_{vs}+\mathbb{E}(\mathbf{q}_b\mathbf{s}^H),\label{eqn047}\\
&=\mathbf{C}_{qs}\label{eqn048}.
\end{IEEEeqnarray}
The derivation from \eqref{eqn047} to \eqref{eqn048} is due to the mutual independence between $\mathbf{s}$ and $\mathbf{v}$. 
Appendix \ref{D} shows 
\begin{equation}\label{eqn049}
\mathbf{C}_{qs}=\mathbf{0}.
\end{equation}
The result \eqref{eqn045} is therefore proved. 

It is perhaps worthwhile to note that in wireless communications, 
$s_n$ is normally centrosymmetric (such as M-PSK and M-QAM) and equally probable.  In this case, it is not hard to find that the 
condition C4) does hold in reality. 
\end{IEEEproof}

In summary, {\em Corollary \ref{cor1}} exhibits the conditions for the quantization noise to be uncorrelated with the thermal noise as well as 
the noiseless part of the received signal. The condition C3) indicates the need for a sufficiently large number of transmit-antennas ($N$). However, 
this does not mean to require a very large $N$ in practice. Let us take an example of $N=8$.  Each element of $\mathbf{r}$ is a 
superposition of $(2N)=16$ independently generated real random-variables, and this can already lead to a reasonable asymptotical result.

{\em Corollary \ref{cor2}} exhibits the auto-covariance matrix of $\mathbf{q}_b$, which is an asymptotical result for $N\rightarrow\infty$.
The exact form of $\mathbf{C}_{qq}$ is very tedious and we do not have the closed-form. Nevertheless, \eqref{eqn045} already provides 
sufficient physical essence for us to conduct the LMMSE analysis. 
 
Finally, {\em Theorem \ref{thm03}} shows that the quantization noise is uncorrelated with the information-bearing symbols. All of 
these results are useful tools to our LMMSE analysis in Section \ref{sec4}.

\section{LMMSE Analysis with The Vector-SOHE Model}\label{sec4}
 The primary aim of this section is to employ the vector-SOHE model \eqref{eqn037}-\eqref{eqn038} to conduct the LMMSE analysis, with which those interesting phenomena observed in the current LMMSE algorithm can be well explained. In addition, a better understanding of the behavior of the current LMMSE algorithm helps to find us an enhanced version particularly for signals with non-constant modulus modulations. 

\subsection{The SOHE-Based LMMSE Analysis}\label{sec4a}
Vector-SOHE is still a linear model. It does not change the classical form of the LMMSE, i.e., $\mathbf{G}^\star=\mathbf{C}_{sy}\mathbf{C}_{yy}^{-1}$ still holds. Despite, the cross-covariance matrix $\mathbf{C}_{sy}$ can now be computed by
\begin{IEEEeqnarray}{ll}
\mathbf{C}_{sy}&=\mathbb{E}\Big(\mathbf{s}(\lambda_b\mathbf{H}\mathbf{s}+\mat{\varepsilon}_b)^H\Big),\label{eqn050}\\
&=\lambda_b\mathbf{C}_{ss}\mathbf{H}^H+\mathbf{C}_{s\varepsilon},\label{eqn051}\\
&=\lambda_b\mathbf{H}^H.\label{eqn052}
\end{IEEEeqnarray}
The derivation from \eqref{eqn051} to \eqref{eqn052} is due to the fact $\mathbf{C}_{s\varepsilon}=\mathbf{0}$ (see {\em Theorem \ref{thm03}}) as well as the assumption that $x_n, \forall n,$ are uncorrelated with respect to $n$ (see the assumption above \eqref{eqn002}). 

The auto-covariance matrix $\mathbf{C}_{yy}$ can be represented by
\begin{equation}
\mathbf{C}_{yy}=\lambda_b^2\mathbf{HH}^H+\mathbf{C}_{\varepsilon\varepsilon},\label{eqn053}
\end{equation}
where 
\begin{IEEEeqnarray}{ll}\label{eqn054}
\mathbf{C}_{\varepsilon\varepsilon}&=\lambda_b^2N_0\mathbf{I}+\mathbf{C}_{qq}+\lambda_b(\mathbf{C}_{qv}+\mathbf{C}_{vq}),\\
&=\lambda_b^2N_0\mathbf{I}+\mathbf{C}_{qq}+2\lambda_b\Re(\mathbf{C}_{qv}).\label{eqn055}
\end{IEEEeqnarray}
Then, the LMMSE formula can be represented by
\begin{equation}\label{eqn056}
\mathbf{G}^\star=\lambda_b^{-1}\mathbf{H}^H(\mathbf{HH}^H+\lambda_b^{-2}\mathbf{C}_{\varepsilon\varepsilon})^{-1}.
\end{equation}
Provided the conditions C2) and C3), \eqref{eqn056} turns into 
(see {\em Corollary \ref{cor1}})
\begin{equation}\label{eqn057}
\mathbf{G}^\star=\lambda_b^{-1}\mathbf{H}^H(\mathbf{HH}^H+N_0\mathbf{I}+\lambda_b^{-2}\mathbf{C}_{qq})^{-1},
\end{equation}
where $\mathbf{C}_{qq}$ can be substituted by \eqref{eqn044} in {\em Corollary \ref{cor2}}. 

\subsection{Comparison between Various LMMSE Formulas}\label{sec4b}
Given that the generalized-AQNM model (see Section \ref{sec2b3}) was only studied for the $1$-bit quantizer, we mainly conduct 
the LMMSE comparison between the SOHE model and the  (modified) AQNM model. As shown in Section \ref{sec2b2}, 
the modified-AQNM model does not give a different LMMSE formula from the AQNM model when the Gaussian quantization noise is assumed. 
Therefore, our study is quickly focused on the comparison with the AQNM model. 

Basically, there are two major differences in their LMMSE forms:

{\em 1)} The SOHE-LMMSE formula has a scaling factor $\lambda_b^{-1}$, which plays the role of equalizing the scalar ambiguity  
inherent in the SOHE model (see \eqref{eqn037}-\eqref{eqn038}). As shown in {\em Theorem \ref{thm01}}, this scalar ambiguity is introduced 
in the low-resolution quantization procedure. It amplifies the signal energy in the digital domain and vanishes with the increase of resolutions. 
This theoretical conclusion coincides well with the phenomenon observed in the literature (e.g. \cite{nguyen2019linear,9144509}). 

{\em 2)} In the AQNM-LMMSE formula \eqref{eqn009}, the impact of the quantization noise is described by the term 
$\mathrm{nondiag}(\rho\mathbf{HH}^H)$. This implies that the quantization noise is modeled as a linear distortion.
However, such is not the case for the SOHE-LMMSE formula. As shown in \eqref{eqn045} and \eqref{eqn057}, the auto-covariance matrix 
$\mathbf{C}_{qq}$ involves the terms $\sigma_r^2$ and $\sigma_r^4$; and higher-order components are approximated in the SOHE model. 
Although \eqref{eqn045} is only an asymptotical and approximate result, it carries a good implication in the sense that the quantization noise 
would introduce non-linear effects to the LMMSE. Due to the modeling mismatch, the AQNM-LMMSE algorithm can suffer additional 
performance degradation. 

Denote $\mathbf{G}^\star_{\eqref{eqn009}}$ and $\mathbf{G}^\star_{\eqref{eqn057}}$ to be the corresponding LMMSE formulas with 
respect to the AQNM and SOHE models. Section \ref{sec2a} indicates that they share the same size, i.e., $(N)\times(K)$. 
Assuming that $\mathbf{G}^\star_{\eqref{eqn009}}$ has the full row rank, we are able to find a $(N)\times(N)$ matrix $\mathbf{\Theta}$ 
fulfilling 
\begin{equation}\label{eqn058}
\mathbf{\Theta}\mathbf{G}^\star_{\eqref{eqn009}}=\mathbf{G}^\star_{\eqref{eqn057}}.
\end{equation}
Denote $(\mathbf{G}^\star_{\eqref{eqn009}})^\dagger$ to be the pseudo inverse of $\mathbf{G}^\star_{\eqref{eqn009}}$. 
The matrix $\mathbf{\Theta}$ can be obtained through
\begin{equation}\label{eqn059}
\mathbf{\Theta}=\mathbf{G}^\star_{\eqref{eqn057}}\Big(\mathbf{G}^\star_{\eqref{eqn009}}\Big)^\dagger.
\end{equation}
 Therefore, the impact of the modeling mismatch inherent in the AQNM-LMMSE can be mitigated through a linear transform. 
Suppose that the matrix $\mathbf{G}^\star_{\eqref{eqn009}}$ has the full row rank. The modeling-mismatch induced 
performance degradation inherent in the AQNM-LMMSE algorithm can be mitigated through the linear transform specified in 
\eqref{eqn058}, where the scaling factor $\lambda_b$ is incorporated in the matrix $\mathbf{\Theta}$. 

\subsection{Enhancement of The AQNM-LMMSE Algorithm}
The SOHE-LMMSE formula describes more explicitly the impact of non-linear distortion in the channel equalization. 
However, the SOHE-LMMSE formula cannot be directly employed for the channel equalization mainly due to two reasons: 
{\em 1)} the auto-covariance matrix $\mathbf{C}_{qq}$ does not have a closed-form in general; and {\em 2)} the scalar 
$\lambda_b$ defined in \eqref{eqn032} comes only from the first-order Hermite kernel. However, other odd-order Hermite 
kernels also contribute to $\lambda_b$. The omission of the third- and higher-order Hermite kernels can make the computation of 
$\lambda_b$ inaccurate. Fortunately, the analysis in \eqref{eqn058} and\eqref{eqn059} show that the SOHE-LMMSE formula can be translated into 
the AQNM-LMMSE formula through a linear transform. In other words, there is a potential to enhance the AQNM-LMMSE algorithm 
by identifying the linear transform $\mathbf{\Theta}$.

Denote $\hat{\mathbf{s}}_{\eqref{eqn057}}\triangleq\mathbf{G}^\star_{\eqref{eqn057}}\mathbf{y}$ 
and $\hat{\mathbf{s}}_{\eqref{eqn009}}\triangleq\mathbf{G}^\star_{\eqref{eqn009}}\mathbf{y}$ to be the outputs of the SOHE-LMMSE 
channel equalizer and the AQNM-LMMSE channel equalizer, respectively. Applying the result \eqref{eqn058}-\eqref{eqn059} yields
\begin{equation}\label{eqn060}
\hat{\mathbf{s}}_{\eqref{eqn009}}=\mathbf{\Theta}^{-1}\hat{\mathbf{s}}_{\eqref{eqn057}}.
\end{equation}
Generally, it is not easy to identify $\mathbf{\Theta}$ and remove it from $\hat{\mathbf{s}}_{\eqref{eqn009}}$.  On the other hand, 
if $\mathbf{G}^\star_{\eqref{eqn057}}$ and $\mathbf{G}^\star_{\eqref{eqn009}}$ are not too different, \eqref{eqn059} implies that 
$\mathbf{\Theta}$ can be considered to be approximately diagonal. In this case, the linear transform reduces to symbol-level scalar ambiguities. 
Assume that the channel-equalized result $\hat{\mathbf{s}}_{\eqref{eqn057}}$ does not have such scalar ambiguities. It is easy to understand 
the scalar ambiguities of $\hat{\mathbf{s}}_{\eqref{eqn009}}$ come from $\lambda_b\mathbf{G}^\star_{\eqref{eqn009}}\mathbf{H}$. In other 
words, we can have the following approximation 
\begin{equation}\label{eqn061}
\mathbf{\Theta}^{-1}\approx\lambda_b\mathbb{D}\Big(\mathbf{G}^\star_{\eqref{eqn009}}\mathbf{H}\Big).
\end{equation}
In \eqref{eqn061}, $\lambda_b$ is the only unknown notation which must be determined. {\em Theorem \ref{thm01}} shows that 
the effect of $\lambda_b$ is the block-level energy amplification, of which the value can be computed using \eqref{appa6}. Finally, we conclude the following form of enhanced LMMSE channel equalizer (eLMMSE)
\begin{equation}\label{eqn063}
\mathbf{G}_e=\frac{1}{\lambda_b}\mathbb{D}\Big(\mathbf{G}^\star_{\eqref{eqn009}}\mathbf{H}\Big)^{-1}\mathbf{G}^\star_{\eqref{eqn009}}.
\end{equation}

\section{Simulation Results and Discussion}\label{sec5}
Computer simulations were carried out to elaborate our theoretical work in Section \ref{sec3} and Section \ref{sec4}. 
Similar to the AQNM models,  the SOHE model cannot be directly evaluated through computer simulations. 
Nevertheless, their features can be indirectly demonstrated through the evaluation of their corresponding LMMSE channel equalizers. 
Given various LMMSE channel equalizers discussed in Section \ref{sec2} and Section \ref{sec4}, it is perhaps useful to provide a brief summary here for the sake of clarification:
 \begin{itemize}
 \item AQNM-LMMSE: this is the LMMSE channel equalizer shown in \eqref{eqn009}. 
 As shown in Section \ref{sec2b2}, the LMMSE channel equalizer \eqref{eqn017} is equivalent to  \eqref{eqn009}; and thus it is not demonstrated in our simulation results.
 \item B-LMMSE: this is the LMMSE channel equalizer shown in \eqref{eqn024}. This channel equalizer is specially designed and optimized for the $1$-bit quantizer. Therefore, it will only be demonstrated in our simulation results for the $1$-bit quantizer.
 \item N-LMMSE: this is the AQNM-LMMSE channel equalizer normalized by the term $\|\mathbf{G}^\star_{\eqref{eqn009}}\mathbf{y}\|$. 
 \item NB-LMMSE: this is the B-LMMSE channel equalizer normalized by the term $\|\mathbf{G}^\star_{\eqref{eqn024}}\mathbf{y}\|$. 
 Both the N-LMMSE and NB-LMMSE channel equalizers have been studied in \cite{7439790,nguyen2019linear,tsefunda}.
\item e-LMMSE: this is the e-LMMSE channel equalizer proposed in \eqref{eqn063}. As shown in Section \ref{sec4}, this e-LMMSE channel equalizer is driven by the SOHE model. 
 \end{itemize}
\begin{figure}[tb]
	\centering
	\includegraphics[scale=0.25]{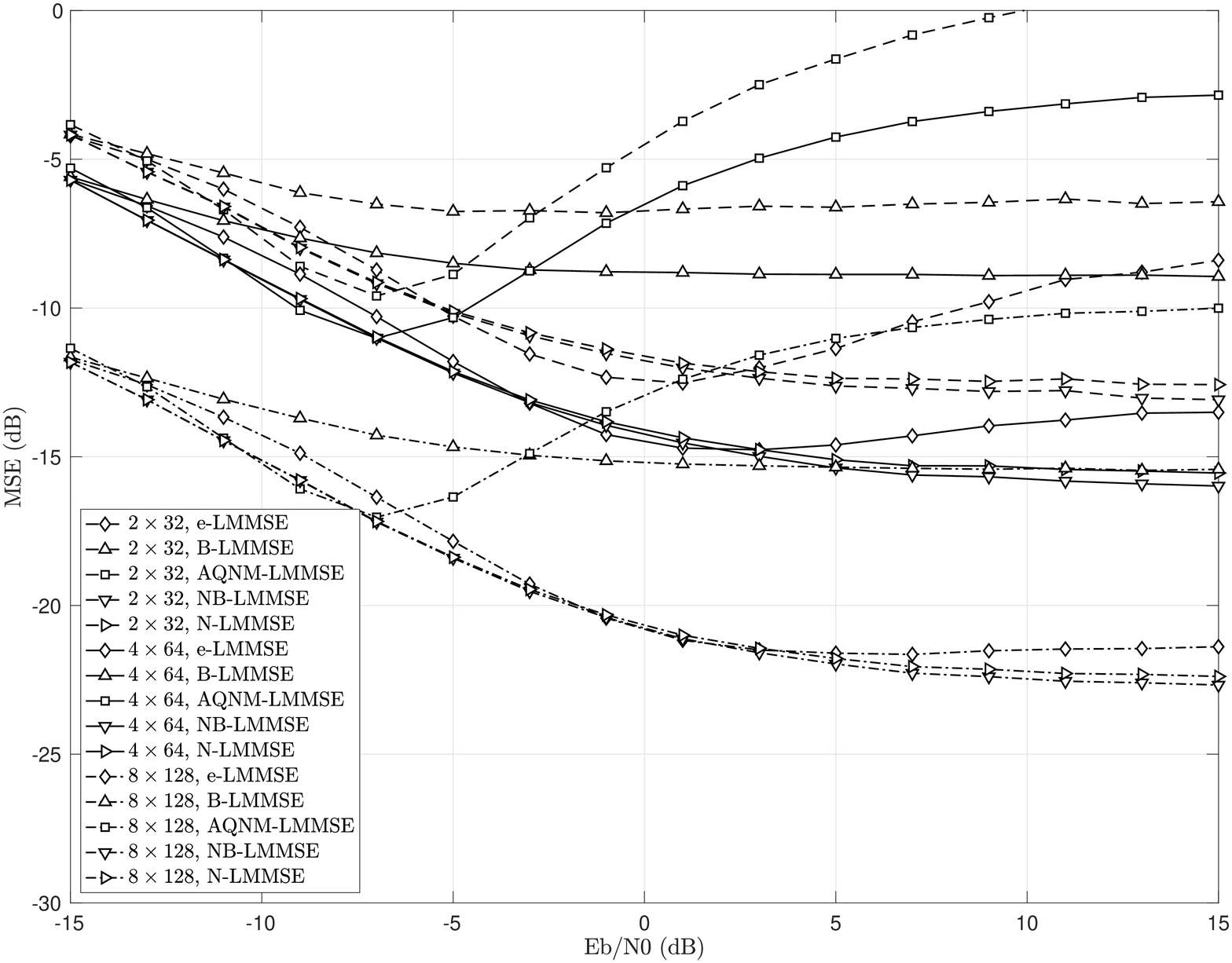}
	\caption{
		The MSE performance as a function of Eb/N0 for the $N$-by-$K$ multiuser-MIMO systems with $1$-bit quantizers, 
		\protect\tikz[baseline]{\protect\draw[line width=0.2mm, dashed] (0,.5ex)--++(0.6,0) ;}~$(N/K)=(2/32)$,
		\protect\tikz[baseline]{\protect\draw[line width=0.2mm] (0,.5ex)--++(0.6,0) ;}~$(N/K)=(4/64)$,
		\protect\tikz[baseline]{\protect\draw[line width=0.2mm, dash dot] (0,.5ex)--++(0.6,0) ;}~$(N/K)=(8/128)$.}\label{fig01}
\end{figure}
In our computer simulations, the e-LMMSE channel equalizer is compared to the SOTA (i.e., AQNM-LMMSE, B-LMMSE, N-LMMSE and NB-LMMSE) in terms of their MSE as well as bit-error-rate (BER) performances. The MSE is defined by 
\begin{equation}\label{eqn064}
\mathrm{MSE}\triangleq\frac{1}{(N)(I)}\sum_{i=0}^{I-1}\|\mathbf{G}_i^\star\mathbf{y}_i-\mathbf{s}_i\|^2,
\end{equation}
where $I$ denotes the number of Monte Carlo trials. 
All the simulation results were obtained by taking average of sufficient number of Monte Carlo trials. For each trial, the wireless MIMO narrowband channel was generated according to independent complex Gaussian distribution (Rayleigh in amplitude), and this is the commonly used simulation setup in the literature \cite{7458830, 6987288}. In addition, the signal-to-noise ratio (SNR) is defined by the average received bit-energy per receive antenna to the noise ratio (Eb/N0), and the transmit power for every transmit antenna is set to be identical. The low-resolution quantization process follows the design in \cite{7037311}, which for 1-bit quantizer, binary quantization is taken; for quantizer other than 1-bit (i.e., 2, 3-bit), the ideal AGC is assumed and the quantization is determined by quantization steps \cite{1057548}.
\begin{figure*}[t]
	\centering
	\includegraphics[scale=0.35]{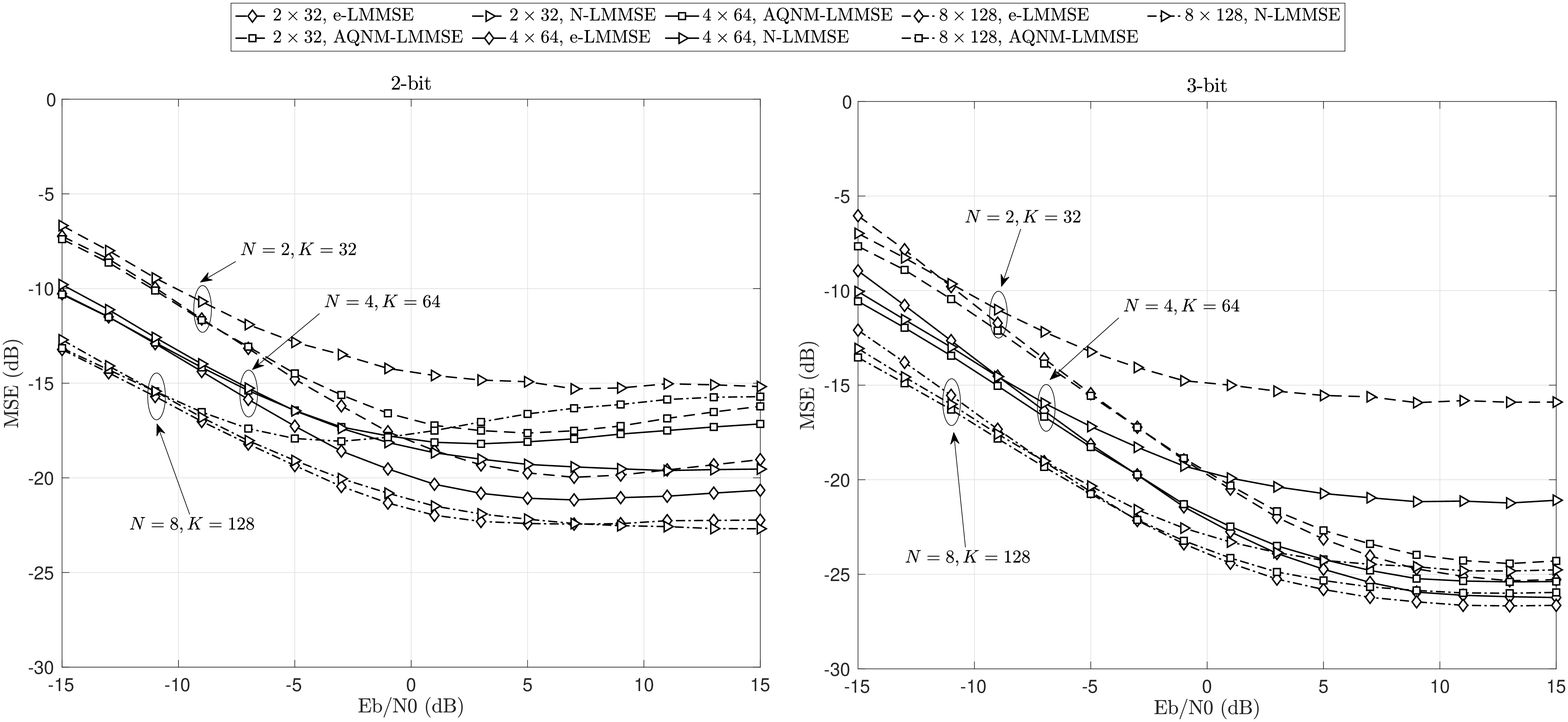}
	\caption{The MSE performance as a function of Eb/N0 for the $N$-by-$K$ multiuser-MIMO systems with $1$-bit quantizers, 
		\protect\tikz[baseline]{\protect\draw[line width=0.2mm, dashed] (0,.5ex)--++(0.6,0) ;}~$(N/K)=(2/32)$,
		\protect\tikz[baseline]{\protect\draw[line width=0.2mm] (0,.5ex)--++(0.6,0) ;}~$(N/K)=(4/64)$,
		\protect\tikz[baseline]{\protect\draw[line width=0.2mm, dash dot] (0,.5ex)--++(0.6,0) ;}~$(N/K)=(8/128)$.}\label{fig02}
\end{figure*}
According to the measures used in computer simulations, we divide the simulation work into two experiments. 
One is designed to examine the MSE performance, and the other is for the BER performance. 
In our simulation results, we demonstrate the performances mainly for $16$-QAM. This is due to two reasons: 
{\em 1)} all types of LMMSE channel equalizers offer the same performances for M-PSK modulations. 
This phenomenon has already been reported in the literature and also discussed in Section \ref{sec1}; 
and {\em 2)} higher-order QAM modulations exhibit almost the same basic features as those of $16$-QAM. 
On the other hand, they perform worse than $16$-QAM due to their increased demand for the resolution of quantizers. 
Those observations are not really novel and thus abbreviated. 

\subsubsection*{Experiment 1}\label{exp1} 

The objective of this experiment is to examine the MSE performance of various LMMSE channel equalizers. 
For all simulations, we keep the transmit antenna to receive antenna ratio to be a constant (e.g. $N/K=1/16$).

\figref{fig01} depicts the MSE performances of various LMMSE channel equalizers as far as the $1$-bit quantizer is concerned. 
Generally, it can be observed that all the MSE performances get improved by increasing the size of MIMO. 
This phenomenon is fully in line with the principle of mMIMO.

It can also be observed that both the AQNM-LMMSE and the B-LMMSE channel equalizers perform poorly throughout the whole SNR range. 
This is because the AQNM models do not capture the scaling ambiguity as described in the SOHE model. When the normalization operation is applied, 
the AQNM-LMMSE and the B-LMMSE channel equalizers turn into their corresponding N-LMMSE and NB-LMMSE equalizers, respectively. 
Interestingly, their performances get significantly improved, and thereby outperforming the e-LMMSE channel equalizer for most of cases. 
On one hand, this is the additional evidence showing the missing of scaling ambiguity in the AQNM models; and on the other hand, 
it is shown that the NB-LMMSE is indeed the optimized LMMSE channel equalizer for the $1$-bit quantizer. 
Nevertheless, we can see that the e-LMMSE approach still offers very comparable MSE performances with the N-LMMSE and NB-LMMSE approaches. 
This provides the indirect evidence showing that the SOHE model offers a good approximation for the $1$-bit quantizer.

Then, we carry on our simulations for $2$- and $3$-bit low-resolution quantizers, respectively, and illustrate their MSE performances in \figref{fig02}. 
It is perhaps worth emphasizing that the B-LMMSE and NB-LMMSE channel equalizers are not examined here since they are devised only for the $1$-bit quantizer. 

The first thing coming into our sight is that the e-LMMSE shows the best MSE performance for almost all the demonstrated cases. 
This is a very good evidence to support our theoretical work about the SOHE model as well as the SOHE-based LMMSE analysis. 
\begin{figure*}[t]
	\centering
	\includegraphics[scale=0.27]{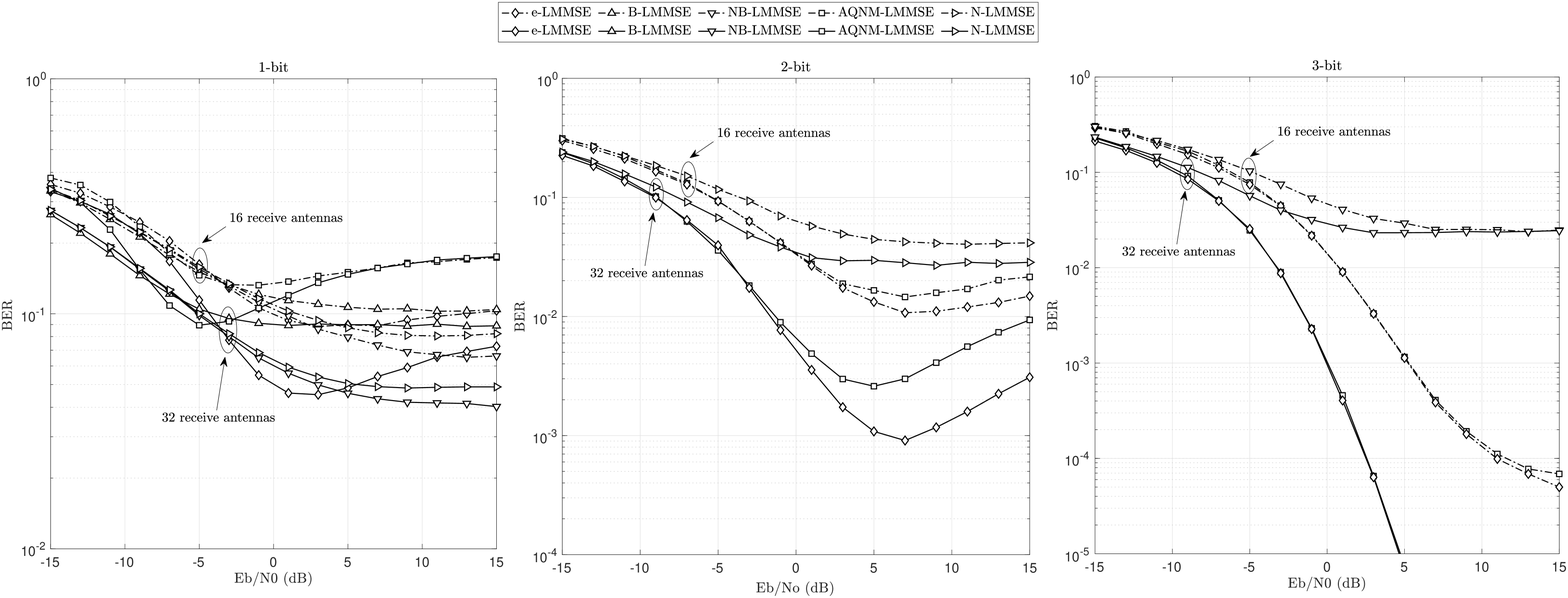}	
	\caption{The BER performance as a function of Eb/N0 for $N= 2$ transmitters, $16$-QAM systems with different resolutions of quantizers, 
		\protect\tikz[baseline]{\protect\draw[line width=0.2mm] (0,.5ex)--++(0.6,0) ;}~$K=32$ receive antennas,
		\protect\tikz[baseline]{\protect\draw[line width=0.2mm, dash dot] (0,.5ex)--++(0.6,0) ;}~$K=16$ receive antennas.}\label{fig03}
\end{figure*}
\begin{figure*}[t]
	\centering
	\includegraphics[scale=0.27]{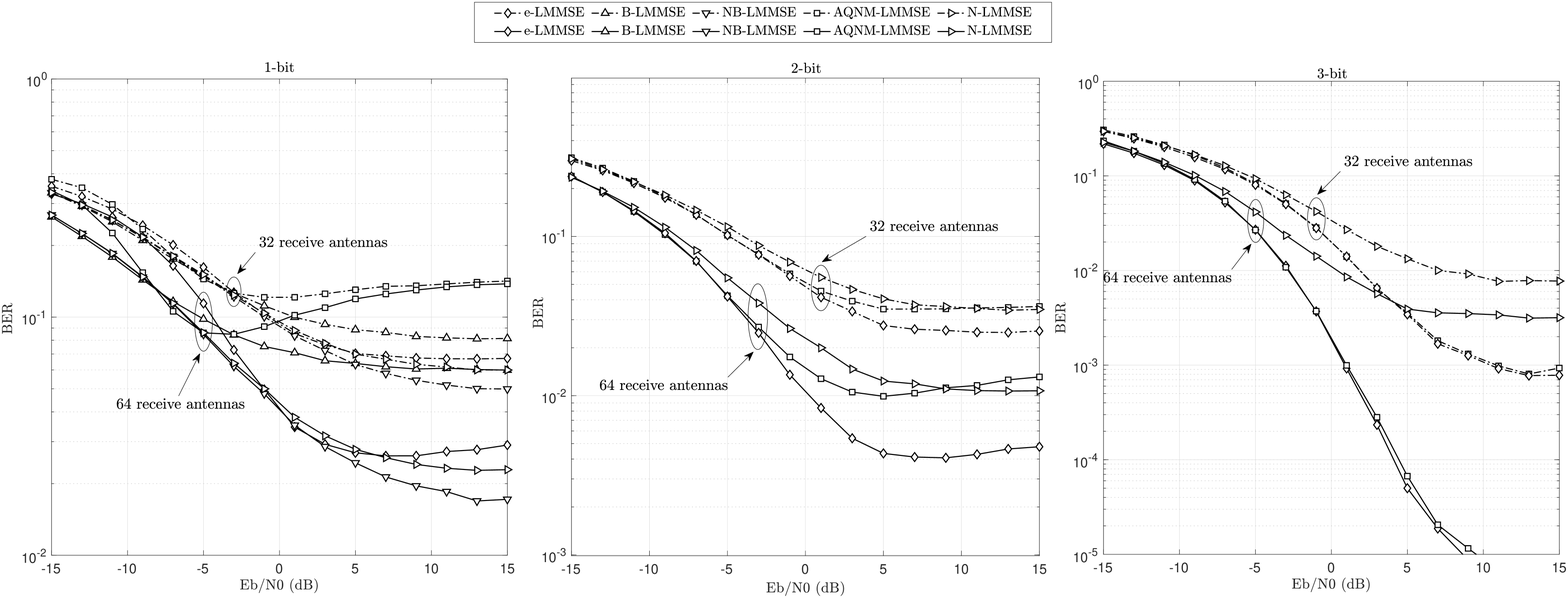}
	\caption{The BER performance as a function of Eb/N0 for $N= 4$ transmitters, $16$-QAM systems with different resolutions of quantizers, 
		\protect\tikz[baseline]{\protect\draw[line width=0.2mm] (0,.5ex)--++(0.6,0) ;}~$K=64$ receive antennas,
		\protect\tikz[baseline]{\protect\draw[line width=0.2mm, dash dot] (0,.5ex)--++(0.6,0) ;}~$K=32$ receive antennas.}\label{fig04}
\end{figure*}

When going down to the detail, specifically for the $2$-bit quantizer, the N-LMMSE approach demonstrates very comparable performance 
with the e-LMMSE approach in the case of larger MIMO (i.e. $(N/K)=(8/128)$). However, its performance gets quickly degraded with the decrease of 
the MIMO size. Take the example of Eb/N0$=5$ dB. For the case of $(N/K)=(8/128)$, 
both the e-LMMSE and the N-LMMSE approaches have their MSEs at around $-22.6$ dB, 
while the AQNM-LMMSE has its MSE at around $-16.8$ dB. Both the e-LMMSE and the N-LMMSE outperform the AQNM-LMMSE by around $6$ dB. 
When the size of MIMO reduces to $(N/K)=(4/64)$, the e-LMMSE shows the best MSE (around $-21.2$ dB). 
The MSE for N-LMMSE and AQNM-LMMSE becomes $-18.9$ dB and $-17.7$ dB, respectively. 
The N-LMMSE underperforms the e-LMMSE by around $2.3$ dB, although it still outperforms the AQNM-LMMSE by around $1.2$ dB. 
By further reducing the size of MIMO to $(N/K)=(2/32)$, the e-LMMSE has its MSE performance degraded to $-19.6$ dB. 
The MSE for N-LMMSE and AQNM-LMMSE now becomes $-14.9$ dB and $-17.4$ dB, respectively. 
The e-LMMSE outperforms the AQNM-LMMSE by around $2.2$ dB and the N-LMMSE by around $4.7$ dB. 
The major reason for this phenomenon to occur is that the AQNM model assumes the quantization distortion and the input signal to be Gaussian.  
This assumption becomes less accurate with the use of less transmit antennas. Moreover, the use of less receive antennas has the spatial de-noising ability reduced. The term used for normalization gets more negatively influenced by the noise as well as the quantization distortion. 
The SOHE model does not assume the input signal and the quantization distortion to be Gaussian, and thus it gets the least negative impact. 
\begin{figure*}[t]
	\centering
	\includegraphics[scale=0.27]{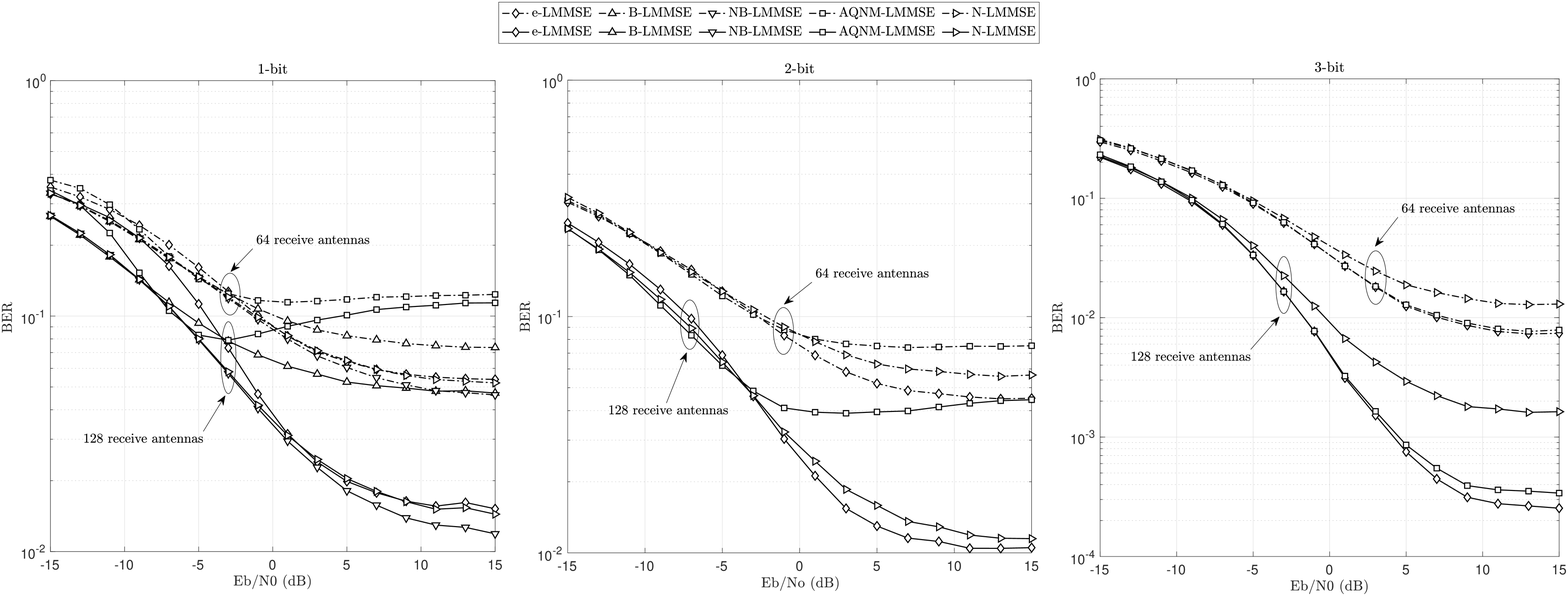}
	
	\caption{The BER performance as a function of Eb/N0 for $N= 8$ transmitters, $16$-QAM systems with different resolutions of quantizers, 
		\protect\tikz[baseline]{\protect\draw[line width=0.2mm] (0,.5ex)--++(0.6,0) ;}~$K=128$ receive antennas,
		\protect\tikz[baseline]{\protect\draw[line width=0.2mm, dash dot] (0,.5ex)--++(0.6,0) ;}~$K=64$ receive antennas.}\label{fig05}
\end{figure*}
Due to the same rationale, the similar phenomenon can also be observed for the $3$-bit quantizer. 
Again, the e-LMMSE approach shows the best performance for almost all the cases. Apart from that, there are two notable differences that worth a mention:
{\em 1)} the performance of AQNM-LMMSE is quite close to that of e-LMMSE for all sizes of MIMO. This is because the $3$-bit quantizer is of reasonably good resolution for $16$-QAM modulations, and this largely mitigates the discrimination between the AQNM model and the SOHE model; 
and 2) the N-LMMSE performs really poorly when compared with the others. This implies the inaccuracy of using the term $\|\mathbf{G}^\star_{\eqref{eqn009}}\mathbf{y}\|$ for the normalization. 

After all, the experiment about the MSE evaluation confirms our theoretical work in Sections \ref{sec2}-\ref{sec4} and demonstrates the major 
advantages of the SOHE model as well as the e-LMMSE channel equalizer from the MSE perspective.

\subsubsection*{Experiment 2}\label{exp2}
It is common knowledge that an MMSE-optimized approach is not necessarily optimized for the detection performance.
This motivates us to examine the average-BER performance for various LMMSE channel equalizers in this experiment.
Basically, this experiment is divided into three sub-tasks, with each having a fixed number of transmit antennas. 

\figref{fig03} depicts the case of $N=2$ transmit antennas. Generally, the use of more receive antennas can largely improve the BER performance. 
This conclusion is true for all types of examined low-resolution quantizers. In other words, all LMMSE channel equalizers can enjoy the 
receiver-side spatial diversity. 

Specifically for the $1$-bit quantizer, the AQNM-based LMMSE  approaches (i.e., AQNM-LMMSE and B-LMMSE) generally underperform their corresponding normalized version (i.e., N-LMMSE and NB-LMMSE). This phenomenon fully coincides with their MSE behaviors shown in 
{\em Experiment 1}- \figref{fig01}. The e-LMMSE approach does not demonstrate remarkable advantages in this special case. It offers the best BER 
at the SNR range around Eb/N0 $=2$ dB, and then the BER grows with the increase of SNR. Such phenomenon is not weird, and this occurs quite 
often in systems with low-resolution quantizers and other non-linear systems due to the physical phenomenon called stochastic resonance \cite{RevModPhys.70.223}. Similar phenomenon also occurs in the AQNM-LMMSE approach. It means that, for low-resolution quantized systems, additive noise could be constructive to the signal detection at certain SNRs, especially for the QAM constellations (e.g. \cite{7247358, 7894211, 9145094, jacobsson2019massive, She2016The}). 
The theoretical analysis of constructive noise in the signal detection can be found in Kay's work \cite{809511}
( interested readers please see Appendix \ref{E} for the elaboration of the phenomenon of constructive noise.)
Interestingly, the normalized approaches do not show 
considerable stochastic resonance phenomenon within the observed SNR range. 
\begin{figure*}[t]
	\centering
	\includegraphics[scale=0.27]{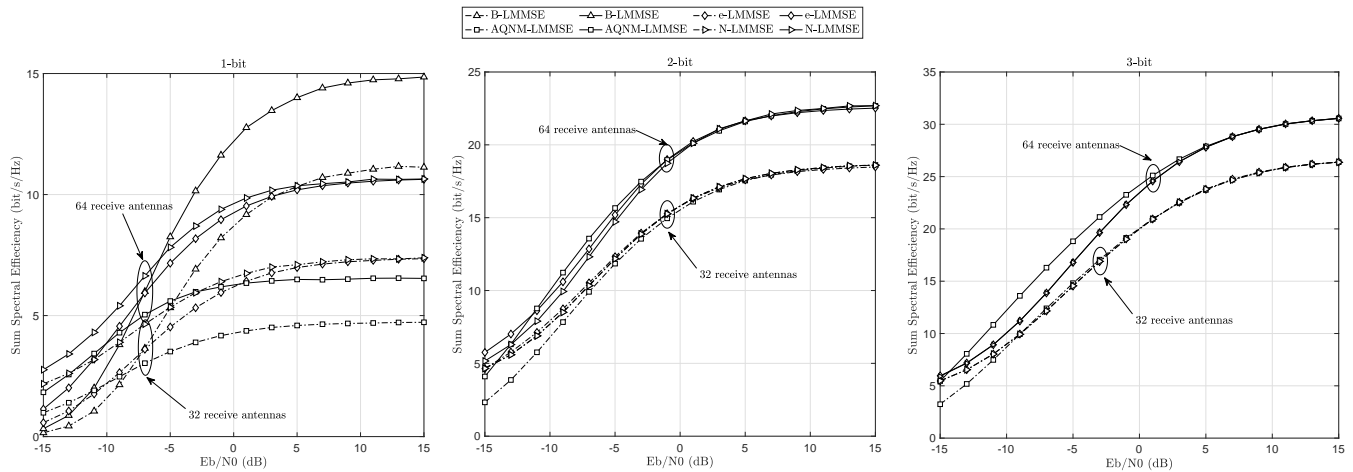}
	
	\caption{The sum SE as a function of Eb/N0 for $N= 4$ transmitters for systems with different resolutions of quantizers, different LMMSE based channel estimators and ZF channel equalizer.
		\protect\tikz[baseline]{\protect\draw[line width=0.2mm] (0,.5ex)--++(0.6,0) ;}~$K=64$ receive antennas,
		\protect\tikz[baseline]{\protect\draw[line width=0.2mm, dash dot] (0,.5ex)--++(0.6,0) ;}~$K=32$ receive antennas.}\label{fig06}
\end{figure*}
When the resolution of quantizer increases to $b=2$ bit, the e-LMMSE approach demonstrates the significant performance gain for most of cases. 
For instance, the e-LMMSE significantly outperforms the AQNM-LMMSE for the higher SNR range (i.e., Eb/N0 $>0$ dB). 
The N-LMMSE approach performs the worst in all the cases. This observation is well in line with our observation in the MSE performance 
(see \figref{fig02}), and they share the same rationale. 

When the resolution of quantizer increases to $b=3$ bit, both the e-LMMSE and the AQNM-LMMSE approaches offer excellent BER performances.
Their performances are very close to each other, and the e-LMMSE only slightly outperforms the AQNM-LMMSE for the case of $K=16$. 
This reason for this phenomenon is the same as that for the MSE performance, which has also been explained in {\em Experiment 1}. 

In a short summary, the e-LMMSE approach shows significant advantages for $2$-bit quantizers. This is the case where the SOHE model offers a 
better mathematical description than the AQNM models, and at the same time the resolution is not high enough to support higher-order modulations. 
This is particularly true for the case of $N=2$ transmit antennas, where the input signal and quantization distortion can be assumed to be white Gaussian.

Now, we increase the number of transmit antennas $(N)$ to investigate how the BER performance will be influenced. 
Accordingly, the number of receive antennas $(K)$ is also increased. The BER results for the case of $N=4$ are plotted in \figref{fig04}. 

Let us begin with the $3$-bit quantizer. We have almost the same observation as for the case of $N=2$ transmit antennas. 
The e-LMMSE approach performs slightly better than the AQNM-LMMSE approach. The performance difference is not really considerable. 
When it comes to the case of $2$-bit quantizer, their difference in BER gets largely increased, and the e-LMMSE approach 
demonstrates significant advantages. It is worth noting that the N-LMMSE approach offers comparable performances 
with the AQNM-LMMSE approach. This is because the increase of transmit antennas brings the input signal and quantization distortion closer to 
the white Gaussian. This rationale is also explained in the MSE analysis. For the case of $1$-bit quantizer, there is no much new phenomenon observed 
in comparison with the case of $N=2$ transmit antennas; apart from that the stochastic resonance phenomenon becomes less significant.

When the number of transmit antennas increases to $N=8$, the BER results are plotted in \figref{fig05}. 
For the case of $3$-bit quantizer, the e-LMMSE approach demonstrates slightly more considerable gain, 
and the N-LMMSE approach gets its performance even closer to the others. 
Similar phenomenon can be observed for the case of $2$-bit quantizer, where the N-LMMSE offers considerably close performance to 
the e-LMMSE approach. The AQNM-LMMSE approach performs the worst. This phenomenon is also observed in the MSE analysis. 
Again, for the $1$-bit quantizer, the NB-LMMSE approach offers the best BER performance, as it is devised and optimized for this special case. 

 Similar to the phenomenon observed in {\em Experiment 1}, the performance of e-LMMSE is not the best for the $1$-bit quantized system. This is because, for the $1$-bit quantized system, there exists an optimum LMMSE channel equalizer using the arcsine law \cite{Mezghani2012,Papoulis_2002}. 
Despite, the proposed e-LMMSE approach can still provide comparable performance over the closed-form approach. When it comes to the $3$-bit quantizer, it can be found that e-LMMSE has only a slight BER gain over the AQNM-LMMSE. It is known that one of the characteristics of SOHE model is that it is not based on the Gaussian quantization noise assumption. However, when the resolution of quantizer rises to $3$-bit, the distribution of quantization noise very approximates to the Gaussian distribution and such results in similar performances between e-LMMSE and AQNM-LMMSE.

\subsubsection*{Experiment 3}\label{exp3}
As response to the reviewers' comments, we add this experiment  to examine the SOHE-based channel estimation and its corresponding channel equalization. 
For this experiment, SOTA approaches can include those results reported in \cite{7931630, 7894211,rao2021massive}.
It is perhaps worth noting that the literature \cite{rao2021massive} considers the use of sigma-delta quantizer, which takes advantage of oversampling to achieve an enhanced performance. 
This is however not the case for our work as well as those in \cite{7931630, 7894211}. 
For the sake of fair comparison, we only conduct the performance comparison between our work and the result in \cite{7931630, 7894211}.
In this experiment, the performance is evaluated through the sum SE defined by \cite{rao2021massive}
\begin{equation}\label{eqn067}
\mathrm{SE} =\frac{T-P}{T}\sum_n^NR_n,
\end{equation}
where $T$ is the length of the coherence interval, and $R_n$ the achievable rate for each transmitter-to-receiver link defined in \cite{7931630, 7894211}. 
This is because the sum SE is the widely considered metric in the SOTA \cite{7931630, 7894211,rao2021massive}, where $T$ is commonly set to $200$.

Similar to \eqref{eqn003}, the mathematical model for low-resolution quantized mMIMO channel estimation is given in the vectorized form
\begin{equation}\label{eqn065}
\mathbf{r}_p = \bar{\mathbf{\Phi}}\bar{\mathbf{h}}+\bar{\mathbf{v}}_p,
\end{equation}
where $\bar{\mathbf{h}}=\mathrm{vec}(\mathbf{H})$, $\bar{\mathbf{\Phi}}=(\mathbf{\Phi} \otimes \mathbf{I}_K)$ and $\mathbf{\Phi}\in \mathbb{C}^{N\times P}$ is the pairwise orthogonal pilot matrix, which is composed of submatrices of the discrete Fourier transform (DFT) operator \cite{Biguesh_1bit}. During training, all $N$ users simultaneously transmit their pilot sequences of $P$ symbols to the BS.
Feeding \eqref{eqn065} to the low-resolution quantizer, we should have the output $\mathbf{y}_p \in \mathbb{C}^{KP\times 1}$, which is similar to \eqref{eqn004}. 
Regarding the LMMSE channel estimation algorithms, we should have the closed-form B-LMMSE estimator for 1-bit quantized model in \cite{7931630} and AQNM-LMMSE and N-LMMSE estimators for other resolutions. 
Those channel estimators are compared with the SOHE-LMMSE channel estimator in \eqref{eqn063}.
Given the LMMSE estimator $\mathbf{W}^*$, the channel estimate can be expressed as $\hat{\mathbf{H}}=\mathrm{unvec}(\mathbf{W}^*\mathbf{y}_b)$. For the sake of fairness, we employ the zero-forcing (ZF) algorithm for the channel equalization as it has been used by the SOTA, i.e., 
$\mathbf{G}_{\text{ZF}} = \mathbf{\hat{\mathbf{H}}}^H(\hat{\mathbf{H}}\hat{\mathbf{H}}^H)^{-1}$.

\figref{fig06} depicts the sum SE performance of various LMMSE channel estimators while $N=4$ transmitters and $K= 32, 64$ receive antennas are considered. The length of the pilot is considered as $P=N$. Similar to the phenomenon observed in above experiments, the rising up of the number of receive antennas and resolution of quantizers can offer significant SE gain.

When the resolution of quantizer is $b=1$ bit, the B-LMMSE algorithm has the best sum SE over other LMMSE channel estimators, and the gap can be approximately 4 bit/s/Hz. This phenomenon is not wired as the B-LMMSE is the closed-form algorithm for 1-bit quantized model \cite{7931630}. SOHE-LMMSE and AQNM-LMMSE channel estimators do not demonstrate advantages in this special scenario, but it can be found that SOHE-LMMSE can achieve almost the same sum SE as the N-LMMSE channel estimator, while AQNM-LMMSE approach performs the worst in such model.

When the resolution of quantizer increases to $b=2$ bit, all three types (i.e., SOHE-LMMSE, AQNM-LMMSE and N-LMMSE) of channel estimators share the similar sum SE. For instance, they can have their sum SE reaching at 16 bit/s/Hz for $K=32$ and 20 bit/s/Hz for $K=64$ for the four-user system. When it comes to the case of 3-bit quantizer, we have almost the same observation as for the case of $b=2$ bit quantizer. The performance difference between all three types of channel estimators is not really considerable for high Eb/N0. When the Eb/N0 $>$ 0dB, for $K=64$, the AQNM-LMMSE channel estimator can slightly outperform the N-LMMSE and SOHE-LMMSE channel estimators. As it is discussed in Section \ref{sec4}-\ref{sec5}, the scalar ambiguity will be detrimental for QAM modulations. However, the setup of each element of the pilot matrix $\mathbf{\Phi}$ follows unit power and all pilot sequences are pairwise orthogonal; similar to the analysis for LMMSE channel equalization for PSK constellations, the scalar ambiguity does not show any side effect on such case. This explains the reason why the SOHE-LMMSE channel estimator has the same sum SE compared with current version LMMSE algorithms. 

\section{Conclusion}
In this paper, a novel linear approximation method, namely SOHE, has been proposed to model the low-resolution quantizers. 
The SOHE model was then extended from the real-scalar form to the complex-vector form, and the latter was applied and extensively studied in 
the low-resolution quantized multiuser-MIMO uplink signal reception. It has been shown that the SOHE model does not require
those assumptions employed in the AQNM model as well as its variations. Instead, it uses the first-order Hermite kernel to model the 
signal part and the second-order Hermite kernel to model the quantization distortion. Such equipped us with sufficient flexibility and 
capacity to develop deeper and novel understanding of the stochastic behavior and correlation characteristics of the quantized signal 
as well as the non-linear distortion. Through our intensive analytical work, it has been unveiled that the low-resolution quantization could result 
in a scalar ambiguity. In the SOHE model, this scalar ambiguity is characterized by the coefficient of the first-order Hermite kernel. 
However, it is not accurately characterized in other linear approximation models due to the white-Gaussian assumption. 
When applying the SOHE model for the LMMSE analysis, 
it has been found that the SOHE-LMMSE formula carries the Hermite coefficient, which equips the SOHE-LMMSE channel equalizer with 
the distinct ability to remove the scalar ambiguity in the channel equalization. It has been shown that the SOHE-LMMSE formula involves 
higher-order correlations, and this prevents the implementation of the SOHE-LMMSE channel equalizer. Nevertheless, it was also found that 
the SOHE-LMMSE formula could be related to the AQNM-LMMSE formula through a certain linear transform. This finding motivated the 
development of the e-LMMSE channel equalizer, which demonstrated significant advantages in the MSE and BER performance evaluation. All of 
the above conclusions have been elaborated through extensive computer simulations in the independent Rayleigh-fading channels. 

\appendices
\section{Proof of  Theorem \ref{thm01}}\label{A}
With the equations \eqref{eqn028} and \eqref{eqn032}, the coefficient $\lambda_b$ can be computed as follows
\begin{IEEEeqnarray}{ll}\label{appa1}
\lambda_b&=\frac{-1}{\sqrt{\pi}}\sum_{m=0}^{M-1}x_m\int_{\tau_m}^{\tau_{m+1}}
\Big[\frac{\partial}{\partial x}\exp(-x^2)\Big]\mathrm{d}x,\\
&=\frac{-1}{\sqrt{\pi}}\sum_{m=0}^{M-1}x_m\int_{\tau_m}^{\tau_{m+1}}(-2x)\exp(-x^2)\mathrm{d}x,\label{appa2}\\
&=\frac{1}{\sqrt{\pi}}\sum_{m=0}^{M-1}x_m\Big(
\exp(-\tau_m^2)-\exp(-\tau_{m+1}^2)
\Big).\label{appa3}
\end{IEEEeqnarray}
We first examine the limit of $\lambda_b$ when $b\rightarrow\infty$. It is equivalent to the following case
\begin{IEEEeqnarray}{ll}
\lim_{b\rightarrow\infty}\lambda_b
&=\frac{1}{\sqrt{\pi}}\lim_{M\rightarrow\infty}\sum_{m=0}^{M-1}x_m\Big(
\exp(-\tau_m^2)\nonumber
\\&\quad\quad\quad\quad\quad\quad\quad\quad\quad\quad-\exp(-\tau_{m+1}^2)\Big). \label{appa4}
\end{IEEEeqnarray}
For $M\rightarrow\infty$, the discrete-time summation in \eqref{appa4} goes back to the integral in \eqref{eqn028}. 
Since it is an ideal quantization, we have $x_m=x$, and thereby having
\begin{equation}\label{appa5}
\lim_{b\rightarrow\infty}\lambda_b=\frac{2}{\sqrt{\pi}}\int_{-\infty}^{\infty}x^2\exp(-x^2)\mathrm{d}x=1.
\end{equation}
The derivation of \eqref{appa5} can be found in \cite[p. 148]{Papoulis_2002}. 

For the symmetric quantization, \eqref{appa3} can be written into
\begin{equation}\label{appa6}
\lambda_b
=\frac{2}{\sqrt{\pi}}\sum_{m=M/2}^{M-1}x_m\Big(
\exp(-\tau_m^2)-\exp(-\tau_{m+1}^2)
\Big).
\end{equation}
Consider the particular range of $x\in(\tau_m, \tau_{m+1}]$ and $\tau_m>0$, in which $\exp(-x^2)$ is a  monotonically 
decreasing function of $x$. Then, we have 
\begin{equation}\label{appa7}
\exp(-\tau_m^2)\geq\exp(-x^2),~x\in(\tau_m, \tau_{m+1}].
\end{equation}
and consequently have
\begin{equation}\label{appa8}
(\tau_{m+1})\exp(-\tau_m^2)\geq\int_0^{\tau_{m+1}}\exp(-x^2)\mathrm{d}x.
\end{equation}
Applying \eqref{eqn034} and \eqref{appa8} into \eqref{appa6} results in
\begin{IEEEeqnarray}{ll}\label{appa9}
\lambda_b&=\frac{2}{\sqrt{\pi}}\sum_{m=M/2}^{M-1}\tau_{m+1}\Big(
\exp(-\tau_m^2)-\exp(-\tau_{m+1}^2)\Big),\\
&\geq\frac{2}{\sqrt{\pi}}\sum_{m=M/2}^{M-1}\Big[\int_0^{\tau_{m+1}}\exp(-x^2)\mathrm{d}x\nonumber\\
&\quad\quad\quad\quad\quad\quad\quad\quad\quad\quad-(\tau_{m+1})\exp(-\tau_{m+1}^2)\Big],\\
&\geq\frac{2}{\sqrt{\pi}}\int_0^{\infty}\exp(-x^2)\mathrm{d}x=1.\label{appa11}
\end{IEEEeqnarray}
{\em Theorem \ref{thm01}} is therefore proved.

\section{Proof of Theorem \ref{thm02}}\label{B}
With the quantization noise model \eqref{eqn033}, the cross-correlation between $x$ and $q_b$ can be computed by
\begin{IEEEeqnarray}{ll}\label{appb1}
\mathbb{E}(xq_b)&\approx\mathbb{E}(x(4\omega_2x^2-2\omega_2)),\\
&\approx4\omega_2\mathbb{E}(x^3)-2\omega_2\mathbb{E}(x).\label{appb2}
\end{IEEEeqnarray}
With the condition C1), \eqref{appb2} is equivalent to 
\begin{equation}\label{appb3}
\mathbb{E}(xq_b)\approx 4\omega_2\mathbb{E}(x^3).
\end{equation}
When $x$ is AWGN, the third-order term $\mathbb{E}(x^3)$ in \eqref{appb3} equals to 0 (see \cite[p. 148]{Papoulis_2002}).
This leads to the observation that $\mathbb{E}(xq_b)=0$
and then the first part of {\em Theorem \ref{thm02}} is therefore proved.

To prove the limit \eqref{eqn036}, we first study the coefficient $\omega_2$ in \eqref{eqn033}. For $b\rightarrow\infty$, 
$\omega_2$ goes back to the formula specified in \eqref{eqn028}. Then, we can compute $\omega_2$ as follows
\begin{IEEEeqnarray}{ll}
\omega_2&=\frac{1}{8\sqrt{\pi}}\int_{-\infty}^{\infty}x\Big[\frac{\partial^2}{\partial x^2}\exp(-x^2)\Big]\mathrm{d}x,\label{appb4}\\
&=\frac{1}{8\sqrt{\pi}}\int_{-\infty}^{\infty}x\Big[(-2+4x^2)\exp(-x^2)\Big]\mathrm{d}x,\label{appb5}\\
&=-\frac{1}{4\sqrt{\pi}}\int_{-\infty}^{\infty}x\exp(-x^2)\mathrm{d}x\nonumber\\
&\quad\quad\quad\quad\quad\quad+\frac{1}{2\sqrt{\pi}}\int_{-\infty}^{\infty}x^3\exp(-x^2)\mathrm{d}x.\label{appb6}
\end{IEEEeqnarray}
It is well known that (also see \cite[p. 148]{Papoulis_2002})
\begin{equation}\label{appb7}
\int_{-\infty}^{\infty}x^l\exp(-x^2)\mathrm{d}x=0, ~l=1, 3;
\end{equation}
and thus we can obtain $\omega_2=0$ for the case of $b\rightarrow\infty$. Applying this result into \eqref{eqn033} leads to 
the conclusion in \eqref{eqn036}. 

\section{Proof of {\em Corollary \ref{cor2}}}\label{C}
With \eqref{eqn039}, we can compute $\mathbf{C}_{qq}$ as follows
\begin{IEEEeqnarray}{ll}
\mathbf{C}_{qq}
&=\mathbb{E}(\mathbf{q}_b\mathbf{q}_b^H),\label{app08}\\
&=4\omega_2^2\Big(4\underbrace{\mathbb{E}\Big(\Re(\mathbf{r})^2+j\Im(\mathbf{r})^2\Big)\Big(\Re(\mathbf{r})^2-j\Im(\mathbf{r})^2\Big)^T}_{\triangleq\mathbf{C}_{qq}^{(1)}}-\nonumber\\
&\quad2\underbrace{\mathbb{E}\Big(\Big(\Re(\mathbf{r})^2+j\Im(\mathbf{r})^2\Big)\otimes\mathbf{1}^T\Big)}_{\triangleq\mathbf{C}_{qq}^{(2)}}-\nonumber\\
&\quad2\underbrace{\mathbb{E}\Big(\mathbf{1}\otimes\Big(\Re(\mathbf{r})^2-j\Im(\mathbf{r})^2\Big)^T\Big)}_{\triangleq\mathbf{C}_{qq}^{(3)}}+\mathbf{1}\otimes\mathbf{1}^T.\Big)
\label{app09}
\end{IEEEeqnarray}
We start from $\mathbf{C}_{qq}^{(2)}$ in \eqref{app09}. Given the conditions C3) and C4), the proof in {\em Corollary \ref{cor1}} shows 
that $\mathbf{r}$ is asymptotically zero-mean complex 
Gaussian with the covariance to be approximately $\sigma_r^2\mathbf{I}$.
\begin{IEEEeqnarray}{ll}
\mathbf{C}_{qq}^{(2)}&=\Big(\mathbb{E}\Big(\Re(\mathbf{r})^2\Big)+j\mathbb{E}\Big(\Im(\mathbf{r})^2\Big)\Big)\otimes\mathbf{1}^T,\label{app10}\\
&=\frac{\sigma_r^2}{2}(\mathbf{1}+j\mathbf{1})\otimes\mathbf{1}^T,\label{app11}
\end{IEEEeqnarray}
Analogously, the following result holds
\begin{equation}
\mathbf{C}_{qq}^{(3)}=\frac{\sigma_r^2}{2}\mathbf{1}\otimes(\mathbf{1}-j\mathbf{1})^T.\label{app12}
\end{equation}
Then, we can obtain
\begin{equation}\label{app13}
2\Big(\mathbf{C}_{qq}^{(2)}+\mathbf{C}_{qq}^{(3)}\Big)=\sigma_r^2\mathbf{1}\otimes\mathbf{1}^T.
\end{equation}
Now, we come to the last term $\mathbf{C}_{qq}^{(1)}$, which can be computed as follows
\begin{IEEEeqnarray}{ll}
\mathbf{C}_{qq}^{(1)}&=\mathbb{E}\Big(\Re(\mathbf{r})^2\Re(\mathbf{r}^T)^2\Big)
+\mathbb{E}\Big(\Im(\mathbf{r})^2\Im(\mathbf{r}^T)^2\Big)+\nonumber\\
&\quad j\Big(\mathbb{E}\Big(\Im(\mathbf{r})^2(\Re(\mathbf{r}^T)^2\Big)-\mathbb{E}\Big(\Re(\mathbf{r})^2(\Im(\mathbf{r}^T)^2\Big)\Big).
\label{app14}
\end{IEEEeqnarray}
Since $\Re(\mathbf{r})$ and $\Im(\mathbf{r})$ follow the identical distribution, we can easily justify
\begin{IEEEeqnarray}{ll}\label{app15}
\mathbb{E}\Big(\Re(\mathbf{r})^2\Re(\mathbf{r}^T)^2\Big)&=\mathbb{E}\Big(\Im(\mathbf{r})^2\Im(\mathbf{r}^T)^2\Big), \\
\mathbb{E}\Big(\Im(\mathbf{r})^2(\Re(\mathbf{r}^T)^2\Big)&=\mathbb{E}\Big(\Re(\mathbf{r})^2(\Im(\mathbf{r}^T)^2\Big).
\label{app16}
\end{IEEEeqnarray}
Applying \eqref{app15} into \eqref{app14} results in
\begin{equation}\label{app17}
\mathbf{C}_{qq}^{(1)}=2\mathbb{E}\Big(\Re(\mathbf{r})^2\Re(\mathbf{r}^T)^2\Big).
\end{equation} 
Plugging \eqref{app17} and \eqref{app13} into \eqref{app09} yields
\begin{equation}\label{app17a}
\mathbf{C}_{qq}=4\omega_2^2\Big(8\mathbb{E}\Big(\Re(\mathbf{r})^2\Re(\mathbf{r}^T)^2\Big)+(1-\sigma_r^2)(\mathbf{1}\otimes\mathbf{1}^T)\Big).
\end{equation}
It is not hard to derive (see \cite[p. 148]{Papoulis_2002})
\begin{equation}\label{app18}
\mathbb{E}\Big(\Re(r_k)^4\Big)=\frac{3\sigma_r^4}{4}.
\end{equation}
\begin{IEEEeqnarray}{ll}
\mathbb{E}\Big(\Re(r_k)^2\Re(r_m)^2\Big)&=\mathbb{E}\Big(\Re(r_k)^2\Big)\mathbb{E}\Big(\Re(r_m)^2\Big), _{\forall k\neq m,}\label{app19}\\
&=\frac{\sigma_r^4}{4}.\label{app20}
\end{IEEEeqnarray}
Applying \eqref{app18} and \eqref{app20} into \eqref{app17} yields
\begin{equation}\label{app21}
\mathbf{C}_{qq}^{(1)}=\frac{\sigma_r^4}{2}(2\mathbf{I}+\mathbf{1}\otimes\mathbf{1}^T).
\end{equation}
Further applying \eqref{app21} into \eqref{app17a} yields the result \eqref{eqn044}. {\em Corollary \ref{cor2}} is therefore proved.

\section{Proof of \eqref{eqn049}}\label{D}
Consider the element-wise cross-correlation between the $m^\mathrm{th}$ element of $\mathbf{q}_b$ (denoted by $q_m$) and the 
$k^\mathrm{th}$ element of $\mathbf{s}$, i.e.,
\begin{IEEEeqnarray}{ll}
\mathbb{E}\Big(q_ms_k^*\Big)&=\mathbb{E}\Big(\Re(s_k)\Re(q_m)+\Im(s_k)\Im(q_m)\Big)+\nonumber\\
&\quad j\mathbb{E}\Big(\Re(s_k)\Im(q_m)-\Im(s_k)\Re(q_m)\Big),\label{app01}\\
&=2\mathbb{E}\Big(\Re(s_k)\Re(q_m)\Big).\label{app02}
\end{IEEEeqnarray}
Using \eqref{eqn033}, we can obtain 
\begin{IEEEeqnarray}{ll}
\mathbb{E}\Big(\Re(s_k)\Re(q_m)\Big)&=\mathbb{E}\Big(\Re(s_k)(4\omega_2\Re(r_m)^2-2\omega_2)\Big),\nonumber\label{app03}\\
&=4\omega_2\mathbb{E}\Big(\Re(s_k)\Re(r_m)^2\Big).\label{app04}
\end{IEEEeqnarray}
The term $\Re(r_m)$ can be represented by
\begin{equation}\label{app05}
\Re(r_m)=\Re(s_k)\Re(h_{m,k})+\gamma_m+\Re(v_m),
\end{equation}
where $\gamma_m$ is the sum of all corresponding terms that are uncorrelated with $\Re(s_k)$, and $h_{m,k}$ is the $(m,k)^\mathrm{th}$
entry of $\mathbf{H}$. Define $\epsilon_m\triangleq\gamma_m+\Re(v_m)$. We apply \eqref{app05} into \eqref{app04} and obtain
\begin{IEEEeqnarray}{ll}
\mathbb{E}&\Big(\Re(s_k)\Re(r_m)^2\Big)=\Re(h_{m,k})^2\mathbb{E}\Big(\Re(s_k)^3\Big)+\nonumber\\
&\quad\quad\underbrace{2\Re(h_{m,k})\mathbb{E}\Big(\Re(s_k)^2\epsilon_m\Big)+\mathbb{E}\Big(\Re(s_k)\epsilon_m^2\Big)}_{=0}.\label{app06}
\end{IEEEeqnarray}
Plugging \eqref{app06} into \eqref{app04} yields
\begin{equation}\label{app07}
\mathbb{E}\Big(\Re(s_k)\Re(q_m)\Big)=4\omega_2\Re(h_{m,k})^2\mathbb{E}\Big(\Re(s_k)^3\Big).
\end{equation}
The condition C4) ensures that the third-order central moments $\mathbb{E}\Big(\Re(s_k)^3\Big)=0$. Hence, we can conclude 
$\mathbb{E}\Big(q_ms_k^*\Big)=0, \forall m,k$. The result \eqref{eqn049} is therefore proved.

\section{Elaborative Explanation of the Phenomenon of Constructive Noise}\label{E}
As response to the review comment, we find it important to elaborate the phenomenon of constructive noise in the low-resolution signal detection. To better explain the phenomenon, we consider the case where two different information-bearing symbol blocks termed $\mathbf{s}^{(1)}$ and $\mathbf{s}^{(2)}$, $\mathbf{s}^{(1)}\neq\mathbf{s}^{(2)}$, are transmitted to the receiver separately. 
In the case of very high SNR or perhaps more extremely the noiseless case, their corresponding received blocks can be expressed by 
\begin{equation}\label{appe1}
\mathbf{r}^{(1)}=\mathbf{H}\mathbf{s}^{(1)},~\mathbf{r}^{(2)}=\mathbf{H}\mathbf{s}^{(2)},
\end{equation}
where the noise $\mathbf{v}$ is omitted for now because it is negligibly small. 
In this linear system, there exists a perfect bijection between $\mathbf{s}$ and $\mathbf{r}$ and we have $\mathbf{r}^{(1)}\neq \mathbf{r}^{(2)}$. 
For this reason, the receiver can reconstruct the information-bearing symbol block from $\mathbf{r}$ without error. 
The noise will only introduce the detrimental impact to the signal detection. 
However, such is not the case for the system with low-resolution ADC.

To make the concept easy to access, we consider the special case of $1$-bit ADC, the output of which is 
\begin{equation}\label{appe2}
\mathbf{y}^{(1)}=\mathcal{Q}_b(\mathbf{H}\mathbf{s}^{(1)}),~\mathbf{y}^{(2)}=\mathcal{Q}_b(\mathbf{H}\mathbf{s}^{(2)}). 
\end{equation}
The nonlinear function $\mathcal{Q}_b(\cdot)$ can destroy the input-output bijection as hold in the linear system. 
Here, we use a simple numerical example to explain the phenomenon. 
To fulfill the condition $\mathbf{s}^{(1)}\neq\mathbf{s}^{(2)}$, we let $\mathbf{s}^{(1)}=[-1+3j, 3-j]^T$ and $\mathbf{s}^{(2)}=[-3+1j, 1-3j]^T$.
Moreover, to simply our discussion, we let $\mathbf{H}=[\mathbf{I}_{2\times2}, \mathbf{I}_{2\times2}]^T$. 
Then, the output of the $1$-bit ADC is given by
\begin{equation}\label{appe3}
\mathbf{y}^{(1)}=\mathbf{y}^{(2)}=[-1+j, 1-j, -1+j, 1-j]^T.
\end{equation}
In the probability domain, we have 
\begin{equation}\label{appe4}
\mathrm{Pr}(\mathbf{y}^{(1)}\neq\mathbf{y}^{(2)}|\mathbf{H}, \mathbf{x}^{(1)}, \mathbf{x}^{(2)})=0.
\end{equation}
It means that there is no bijection between $\mathbf{y}$ and $\mathbf{s}$ in this case; and for this reason, the receiver is not able to successfully reconstruct $\mathbf{s}$ from $\mathbf{y}$ even in the noiseless case. 

Now, we increase the noise power (or equivalently reduce the SNR). 
Due to the increased randomness, we understand that a positive-amplitude signal can become a negative-amplitude signal. 
Denote $s$ to be a real scalar drawn from the discrete finite-set $\{-3, -1, 1, 3\}$ and $v$ the Gaussian noise. It is trivial to have
\begin{equation}\label{appe5}
\mathrm{Pr}(s+v>0|s=-1)>\mathrm{Pr}(s+v>0|s=-3).
\end{equation}
As shown in \cite{9145094}, with the decrease of SNR from a large value (e.g., the noiseless case), the difference between these two probabilities will quickly increase at the beginning, and then converge to a certain non-zero value. 
It means that the noise helps to discriminate the ADC output $\mathbf{y}^{(1)}$ and $\mathbf{y}^{(2)}$, i.e.,
\begin{equation}\label{appe6}
\mathrm{Pr}(\mathbf{y}^{(1)}\neq\mathbf{y}^{(2)}|\mathbf{H}, \mathbf{x}^{(1)}, \mathbf{x}^{(2)})\neq 0,
\end{equation}
and the probability increases with the decrease of SNR, and this helps the signal detectability \cite{809511}. 
Since the probability converges to a certain value at some SNR, further reducing the SNR will not improve the signal detectability but will only degrade the detection performance. For the general case, the converged probability of \eqref{appe6} can be found in \cite{9145094}, i.e.,
\begin{equation}\label{rev01}
	\mathrm{Pr}(\mathbf{y}^{(1)}=\mathbf{y}^{(2)}|\mathbf{s}^{(1)}\neq\mathbf{s}^{(2)})
	=\frac{(\mathcal{L}^N)(\mathcal{L}^N-1)}{2^{(2K+1)}},
\end{equation} 
where $\mathcal{L}$ is the modulation order. Finally, when the resolution of the quantizer increases, the communication system becomes closer to linear, for which the noise becomes less constructive.

\ifCLASSOPTIONcaptionsoff
  \newpage
\fi

\bibliographystyle{IEEEtran}
\bibliography{ref}	

\begin{thebibliography}{10}
\providecommand{\url}[1]{#1}
\csname url@samestyle\endcsname
\providecommand{\newblock}{\relax}
\providecommand{\bibinfo}[2]{#2}
\providecommand{\BIBentrySTDinterwordspacing}{\spaceskip=0pt\relax}
\providecommand{\BIBentryALTinterwordstretchfactor}{4}
\providecommand{\BIBentryALTinterwordspacing}{\spaceskip=\fontdimen2\font plus
\BIBentryALTinterwordstretchfactor\fontdimen3\font minus
  \fontdimen4\font\relax}
\providecommand{\BIBforeignlanguage}[2]{{%
\expandafter\ifx\csname l@#1\endcsname\relax
\typeout{** WARNING: IEEEtran.bst: No hyphenation pattern has been}%
\typeout{** loaded for the language `#1'. Using the pattern for}%
\typeout{** the default language instead.}%
\else
\language=\csname l@#1\endcsname
\fi
#2}}
\providecommand{\BIBdecl}{\relax}
\BIBdecl

\bibitem{6375940}
F.~{Rusek}, D.~{Persson}, B.~K. {Lau}, E.~G. {Larsson}, T.~L. {Marzetta},
  O.~{Edfors}, and F.~{Tufvesson}, ``Scaling up {MIMO}: Opportunities and
  challenges with very large arrays,'' \emph{IEEE Signal Process. Mag.},
  vol.~30, no.~1, pp. 40--60, Dec. 2013.

\bibitem{1519678}
{X. Zhang}, A.~{Molisch}, and {S.-Y. Kung}, ``Variable-phase-shift-based
  {RF}-baseband codesign for {MIMO} antenna selection,'' \emph{IEEE Trans.
  Signal Process.}, vol.~53, no.~11, pp. 4091--4103, Nov. 2005.

\bibitem{6457363}
H.~Q. {Ngo}, E.~G. {Larsson}, and T.~L. {Marzetta}, ``Energy and spectral
  efficiency of very large multiuser {MIMO} systems,'' \emph{IEEE Trans.
  Commun.}, vol.~61, no.~4, pp. 1436--1449, Apr. 2013.

\bibitem{761034}
R.~H. Walden, ``Analog-to-digital converter survey and analysis,'' \emph{IEEE
  J. Sel. Areas Commun.}, vol.~17, no.~4, pp. 539--550, Apr. 1999.

\bibitem{5351659}
J.~{Singh}, O.~{Dabeer}, and U.~{Madhow}, ``On the limits of communication with
  low-precision analog-to-digital conversion at the receiver,'' \emph{IEEE
  Trans. Commun.}, vol.~57, no.~12, pp. 3629--3639, Dec. 2009.

\bibitem{DBLP:journals/corr/RisiPL14}
\BIBentryALTinterwordspacing
C.~Risi, D.~Persson, and E.~G. Larsson, ``Massive {MIMO} with 1-bit {ADC},''
  \emph{CoRR}, vol. abs/1404.7736, 2014. [Online]. Available:
  \url{http://arxiv.org/abs/1404.7736}
\BIBentrySTDinterwordspacing

\bibitem{6891254}
E.~{Björnson}, J.~{Hoydis}, M.~{Kountouris}, and M.~{Debbah}, ``Massive {MIMO}
  systems with non-ideal hardware: Energy efficiency, estimation, and capacity
  limits,'' \emph{IEEE Trans. Inf. Theory}, vol.~60, no.~11, pp. 7112--7139,
  Nov. 2014.

\bibitem{7106472}
X.~{Zhang}, M.~{Matthaiou}, M.~{Coldrey}, and E.~{Björnson}, ``Impact of
  residual transmit {RF} impairments on training-based {MIMO} systems,''
  \emph{IEEE Trans. Commun.}, vol.~63, no.~8, pp. 2899--2911, Aug. 2015.

\bibitem{7080890}
E.~{Björnson}, M.~{Matthaiou}, and M.~{Debbah}, ``Massive {MIMO} with
  non-ideal arbitrary arrays: Hardware scaling laws and circuit-aware design,''
  \emph{IEEE Trans. Wireless Commun.}, vol.~14, no.~8, pp. 4353--4368, Aug.
  2015.

\bibitem{6987288}
S.~{Wang}, Y.~{Li}, and J.~{Wang}, ``Multiuser detection in massive spatial
  modulation {MIMO} with low-resolution {ADC}s,'' \emph{IEEE Trans. Wireless
  Commun.}, vol.~14, no.~4, pp. 2156--2168, Apr. 2015.

\bibitem{7088639}
J.~{Choi}, D.~J. {Love}, D.~R. {Brown}, and M.~{Boutin}, ``Quantized
  distributed reception for {MIMO} wireless systems using spatial
  multiplexing,'' \emph{IEEE Trans. Signal Process.}, vol.~63, no.~13, pp.
  3537--3548, Jul. 2015.

\bibitem{9311778}
M.~Shao and W.-K. Ma, ``Binary {MIMO} detection via homotopy optimization and
  its deep adaptation,'' \emph{IEEE Trans. Signal Process.}, vol.~69, pp.
  781--796, Feb. 2021.

\bibitem{4475570}
A.~{Mezghani}, M.~{Khoufi}, and J.~A. {Nossek}, ``Maximum likelihood detection
  for quantized {MIMO} systems,'' in \emph{Proc. Int. ITG Workshop Smart
  Antennas}, Feb. 2008, pp. 278--284.

\bibitem{9145094}
L.~{Liu}, Y.~{Ma}, and R.~{Tafazolli}, ``{MIMO} or {SIMO} for wireless
  communications with binary-array receivers,'' in \emph{Proc. IEEE Int. Conf.
  Commun. Workshop (ICCW)}, Jun. 2020, pp. 1--6.

\bibitem{7439790}
J.~Choi, J.~Mo, and R.~W. Heath, ``Near maximum-likelihood detector and channel
  estimator for uplink multiuser massive {MIMO} systems with one-bit {ADC}s,''
  \emph{IEEE Trans. Commun.}, vol.~64, no.~5, pp. 2005--2018, May 2016.

\bibitem{8240630}
S.~{Hong}, S.~{Kim}, and N.~{Lee}, ``A weighted minimum distance decoding for
  uplink multiuser {MIMO} systems with low-resolution {ADC}s,'' \emph{IEEE
  Trans. Commun.}, vol.~66, no.~5, pp. 1912--1924, May 2018.

\bibitem{8345169}
Y.~{Jeon}, N.~{Lee}, S.~{Hong}, and R.~W. {Heath}, ``One-bit sphere decoding
  for uplink massive {MIMO} systems with one-bit {ADCs},'' \emph{IEEE Trans.
  Wireless Commun.}, vol.~17, no.~7, pp. 4509--4521, Jul. 2018.

\bibitem{7355388}
C.~{Wen}, C.~{Wang}, S.~{Jin}, K.~{Wong}, and P.~{Ting}, ``Bayes-optimal joint
  channel-and-data estimation for massive {MIMO} with low-precision {ADC}s,''
  \emph{IEEE Trans. Signal Process.}, vol.~64, no.~10, pp. 2541--2556, May
  2016.

\bibitem{7426735}
J.~T. Parker and P.~Schniter, ``Parametric bilinear generalized approximate
  message passing,'' \emph{IEEE J. Sel Topics Signal Process.}, vol.~10, no.~4,
  pp. 795--808, Jun. 2016.

\bibitem{8234637}
Z.~Zhang, X.~Cai, C.~Li, C.~Zhong, and H.~Dai, ``One-bit quantized massive
  {MIMO} detection based on variational approximate message passing,''
  \emph{IEEE Trans. Signal Process.}, vol.~66, no.~9, pp. 2358--2373, May 2018.

\bibitem{mezghani11}
A.~Mezghani, M.~Rouatbi, and J.~A. Nossek, ``A modified {MMSE} receiver for
  quantized {MIMO} systems,'' in \emph{Proc. Int. ITG Workshop Smart Antennas},
  Feb. 2007, pp. 1--5.

\bibitem{7307134}
L.~{Fan}, S.~{Jin}, C.~{Wen}, and H.~{Zhang}, ``Uplink achievable rate for
  massive {MIMO} systems with low-resolution {ADC},'' \emph{IEEE Commun.
  Lett.}, vol.~19, no.~12, pp. 2186--2189, Dec. 2015.

\bibitem{7308988}
O.~{Orhan}, E.~{Erkip}, and S.~{Rangan}, ``Low power analog-to-digital
  conversion in millimeter wave systems: Impact of resolution and bandwidth on
  performance,'' in \emph{Proc. IEEE Inf. Theory Appl. Workshop (ITAW)}, Dec.
  2015, pp. 191--198.

\bibitem{7876856}
J.~{Mo}, A.~{Alkhateeb}, S.~{Abu-Surra}, and R.~W. {Heath}, ``Hybrid
  architectures with few-bit {ADC} receivers: Achievable rates and energy-rate
  tradeoffs,'' \emph{IEEE Trans. Wireless Commun.}, vol.~16, no.~4, pp.
  2274--2287, Apr. 2017.

\bibitem{7896590}
Y.~{Dong} and L.~{Qiu}, ``Spectral efficiency of massive {MIMO} systems with
  low-resolution {ADC}s and {MMSE} receiver,'' \emph{IEEE Commun. Lett},
  vol.~21, no.~8, pp. 1771--1774, Aug. 2017.

\bibitem{7420605}
J.~{Zhang}, L.~{Dai}, S.~{Sun}, and Z.~{Wang}, ``On the spectral efficiency of
  massive {MIMO} systems with low-resolution {ADC}s,'' \emph{IEEE Commun.
  Lett.}, vol.~20, no.~5, pp. 842--845, May 2016.

\bibitem{Mezghani2012}
A.~Mezghani and J.~Nossek, ``Capacity lower bound of {MIMO} channels with
  output quantization and correlated noise,'' in \emph{Proc. IEEE Int. Symp.
  Inf. Theory (ISIT)}, Jul. 2012.

\bibitem{Bussgang52}
J.~J. {Bussgang}, ``Crosscorrelation functions of amplitude-distorted
  {Gaussian} signals,'' \emph{Res. Lab. Electron., Massachusetts Inst. Techno.,
  Cambridge, MA, USA, Tech. Rep. 216}, Mar. 1952.

\bibitem{nguyen2019linear}
\BIBentryALTinterwordspacing
L.~V. Nguyen and D.~H.~N. Nguyen, ``Linear receivers for massive {MIMO} systems
  with one-bit {ADC}s,'' 2019. [Online]. Available:
  \url{https://arxiv.org/abs/1907.06664.}
\BIBentrySTDinterwordspacing

\bibitem{7931630}
Y.~{Li}, C.~{Tao}, G.~{Seco-Granados}, A.~{Mezghani}, A.~L. {Swindlehurst}, and
  L.~{Liu}, ``Channel estimation and performance analysis of one-bit massive
  {MIMO} systems,'' \emph{IEEE Trans. Signal Process.}, vol.~65, no.~15, pp.
  4075--4089, Aug. 2017.

\bibitem{7894211}
S.~{Jacobsson}, G.~{Durisi}, M.~{Coldrey}, U.~{Gustavsson}, and C.~{Studer},
  ``Throughput analysis of massive {MIMO} uplink with low-resolution {ADC}s,''
  \emph{IEEE Trans. Wireless Commun.}, vol.~16, no.~6, pp. 4038--4051, Jun.
  2017.

\bibitem{8337813}
J.~{Zhang}, L.~{Dai}, X.~{Li}, Y.~{Liu}, and L.~{Hanzo}, ``On low-resolution
  {ADC}s in practical {5G} millimeter-wave massive {MIMO} systems,'' \emph{IEEE
  Commun. Mag.}, vol.~56, no.~7, pp. 205--211, Jul. 2018.

\bibitem{tsefunda}
D.~Tse and P.~Viswanath, \emph{Fundamentals of Wireless Communication}.\hskip
  1em plus 0.5em minus 0.4em\relax USA: Cambridge University Press, 2005.

\bibitem{9144509}
S.~{Gayan}, R.~{Senanayake}, H.~{Inaltekin}, and J.~{Evans}, ``Low-resolution
  quantization in phase modulated systems: Optimum detectors and error rate
  analysis,'' \emph{IEEE Open J. Commun. Society}, vol.~1, pp. 1000--1021, Aug.
  2020.

\bibitem{7247358}
S.~{Jacobsson}, G.~{Durisi}, M.~{Coldrey}, U.~{Gustavsson}, and C.~{Studer},
  ``One-bit massive {MIMO}: Channel estimation and high-order modulations,'' in
  \emph{Proc. IEEE Int. Conf. Commun. Workshop (ICCW)}, Jun. 2015, pp.
  1304--1309.

\bibitem{Proakis2007}
J.~Proakis, \emph{Digital Communications. 5th Edition}.\hskip 1em plus 0.5em
  minus 0.4em\relax McGraw Hill, 2007.

\bibitem{5592653}
S.~{Krone} and G.~{Fettweis}, ``Fading channels with 1-bit output quantization:
  Optimal modulation, ergodic capacity and outage probability,'' in \emph{Proc.
  IEEE Inf. Theory Workshop (ITW)}, 2010, pp. 1--5.

\bibitem{8320852}
H.~{He}, C.~{Wen}, and S.~{Jin}, ``Bayesian optimal data detector for hybrid
  {mmWave} {MIMO-OFDM} systems with low-resolution {ADC}s,'' \emph{IEEE J. Sel.
  Topics Signal Process.}, vol.~12, no.~3, pp. 469--483, Jun. 2018.

\bibitem{8610159}
Y.~{Jeon}, H.~{Do}, S.~{Hong}, and N.~{Lee}, ``Soft-output detection methods
  for sparse millimeter-{Wave} {MIMO} systems with low-precision {ADC}s,''
  \emph{IEEE Trans. Commun.}, vol.~67, no.~4, pp. 2822--2836, Apr. 2019.

\bibitem{7155570}
J.~Mo and R.~W. Heath, ``Capacity analysis of one-bit quantized {MIMO} systems
  with transmitter channel state information,'' \emph{IEEE Trans. Signal
  Process.}, vol.~63, no.~20, pp. 5498--5512, Oct. 2015.

\bibitem{5501995}
O.~{Dabeer} and U.~{Madhow}, ``Channel estimation with low-precision
  analog-to-digital conversion,'' in \emph{Proc. IEEE Int. Conf. Commun.
  (ICC)}, Jun. 2010, pp. 1--6.

\bibitem{708938}
T.~M. {Lok} and V.~K.-W. {Wei}, ``Channel estimation with quantized
  observations,'' in \emph{Proc. IEEE Int. Symp. Inf. Theory (ISIT)}, Aug.
  1998, p. 333.

\bibitem{1057548}
J.~{Max}, ``Quantizing for minimum distortion,'' \emph{IRE Trans. Inf. Theory},
  vol.~6, no.~1, pp. 7--12, Mar. 1960.

\bibitem{664234}
J.~Khoury, ``On the design of constant settling time {AGC} circuits,''
  \emph{IEEE Trans. Circuits Syst. II: Analog Digit. Signal Process.}, vol.~45,
  no.~3, pp. 283--294, Mar. 1998.

\bibitem{1092057}
J.~Ohlson, ``Exact dynamics of automatic gain control,'' \emph{IEEE Trans.
  Commun.}, vol.~22, no.~1, pp. 72--75, Jan. 1974.

\bibitem{Liu2021vtc}
L.~Liu, S.~Xue, Y.~Ma, N.~Yi, and R.~Tafazolli, ``On the design of quantization
  functions for uplink massive {MIMO} with low-resolution {ADCs},'' in
  \emph{Proc. IEEE Veh. Techno. Conf. (VTC-Spring)}, Apr. 2021, pp. 1--5.

\bibitem{Poularikas_1999}
A.~D. {Poularikas}, \emph{Handbook of Formulas and Tables for Signal
  Processing}, ser. Electron. Eng. Handbook.\hskip 1em plus 0.5em minus
  0.4em\relax Springer Berlin Heidelberg, 1998.

\bibitem{60086}
J.-B. {Martens}, ``The {Hermite} transform-theory,'' \emph{IEEE Trans. Acoust.,
  Speech, Signal Process.}, vol.~38, no.~9, pp. 1595--1606, Sep. 1990.

\bibitem{7458830}
C.~{Studer} and G.~{Durisi}, ``Quantized massive {MU-MIMO-OFDM} uplink,''
  \emph{IEEE Trans. Commun.}, vol.~64, no.~6, pp. 2387--2399, Jun. 2016.

\bibitem{7037311}
S.~{Wang}, Y.~{Li}, J.~{Wang}, and X.~{Xu}, ``Multiuser detection in massive
  spatial modulation {(SM-) MIMO} with low-resolution {ADC}s,'' in \emph{Proc.
  IEEE Global Commun. Conf. (GLOBECOM)}, Dec. 2014, pp. 3273--3278.

\bibitem{RevModPhys.70.223}
\BIBentryALTinterwordspacing
L.~Gammaitoni, P.~H\"anggi, P.~Jung, and F.~Marchesoni, ``Stochastic
  resonance,'' \emph{Rev. Mod. Phys.}, vol.~70, pp. 223--287, Jan. 1998.
  [Online]. Available: \url{https://link.aps.org/doi/10.1103/RevModPhys.70.223}
\BIBentrySTDinterwordspacing

\bibitem{jacobsson2019massive}
\BIBentryALTinterwordspacing
S.~Jacobsson, L.~Aabel, M.~Coldrey, I.~C. Sezgin, C.~Fager, G.~Durisi, and
  C.~Studer, ``Massive {MU-MIMO-OFDM} uplink with direct {RF}-sampling and
  1-bit {ADC}s,'' 2019. [Online]. Available:
  \url{https://arxiv.org/abs/1907.07091.}
\BIBentrySTDinterwordspacing

\bibitem{She2016The}
C.~She, ``The secret of using noise to improve your {ADC}’s performance -
  analog - technical articles - {TI E2E} support forums,'' 2016.

\bibitem{809511}
S.~{Kay}, ``Can detectability be improved by adding noise?'' \emph{IEEE Signal
  Process. Lett.}, vol.~7, no.~1, pp. 8--10, Jan. 2000.

\bibitem{Papoulis_2002}
A.~Papoulis and S.~U. Pillai, \emph{Probability, Random Variables, and
  Stochastic Processes. 4th Edition}.\hskip 1em plus 0.5em minus 0.4em\relax
  McGraw-Hill Higher Education, 2002.

\bibitem{rao2021massive}
\BIBentryALTinterwordspacing
S.~Rao, G.~Seco-Granados, H.~Pirzadeh, J.~A. Nossek, and A.~L. Swindlehurst,
  ``Massive {MIMO} channel estimation with low-resolution spatial sigma-delta
  {ADC}s,'' 2021. [Online]. Available: \url{https://arxiv.org/abs/2005.07752.}
\BIBentrySTDinterwordspacing

\bibitem{Biguesh_1bit}
M.~Biguesh and A.~B. Gershman, ``Downlink channel estimation in cellular
  systems with antenna arrays at base stations using channel probing with
  feedback,'' \emph{EURASIP J. Appl. Signal Process}, vol. 2004, p.
  1330–1339, Aug. 2004.

\end{thebibliography}

\end{document}